\documentclass[paper=a4, 11pt, DIV=14,colorlinks=true,allcolors=blue,breaklinks=true]{scrartcl}
\usepackage{hyperref}
\usepackage{graphicx}
\usepackage{xspace}
\usepackage{amsmath}
\usepackage{dcolumn}
\usepackage[scientific-notation = fixed, fixed-exponent = 0,%
						round-mode = places, round-precision = 4,%
						retain-explicit-plus]{siunitx}
\usepackage{collcell}
\usepackage{comment}
\usepackage{natbib}
\usepackage{enumitem}
\usepackage{subfigure}
\usepackage{breakcites}

\newcommand{\mobitopp}{mobiTopp\xspace}
\newcommand{\gnuR}{\emph{GNU~R\xspace}}

\newcommand{\stageinit}{long-term model\xspace}
\newcommand{\stagesimu}{short-term model\xspace}

\newcommand{\state}[1]{\emph{#1}}

\newcommand{\stateUninitialized}{\state{not yet initialized}\xspace}
\newcommand{\stateActivity}{\state{execute activity}\xspace}
\newcommand{\stateTrip}{\state{make trip}\xspace}
\newcommand{\stateFinished}{\state{finished}\xspace}
\newcommand{\stateWait}{\state{wait for ride}\xspace}

\newcommand{\activityprogram}{activity schedule\xspace}

\newcommand{\tabindent}{\hspace{2em}}

\newcolumntype{Q}{>{\collectcell\num}r<{\endcollectcell}}
\newcolumntype{q}{>{\sisetup{scientific-notation=fixed,fixed-exponent=0,round-mode=off}}S}
\newcommand{\esc}[1]{\multicolumn{1}{@{}c@{}}{#1}}
\newcommand{\escl}[1]{\multicolumn{1}{@{}l@{}}{#1}}

\newcommand{\institute}[1]{}

\begin{document}

\title{\LARGE Modeling travel demand over a period of one week:\\ The \mobitopp model}

\pagestyle{myheadings}
\markright{Modeling travel demand over a period of one week: The \mobitopp model} 

\author{Nicolai Mallig         \and
        Peter Vortisch 
}
\date{\footnotesize 
              Karlsruhe Institute of Technology \\
              Institute for Transport Studies \\
              Kaiserstrasse 12 \\
              76131 Karlsruhe \\
}

\maketitle

\begin{abstract}

When mobiTopp was initially designed, more than 10 years ago, 
it has been the first 
travel demand simulation model 
intended for an analysis period of one week.
However, the first version supported only an analysis period of one day.
This paper describes the lessons learned while extending the simulation period from one day to one week.
One important issue is ensuring realistic start times of activities. 
Due to differences between the realized trip durations during the simulation and 
the trip durations assumed when creating the activity schedule,
the realized activity schedule and the planned activity schedule 
may deviate from each other at some point in time during simulation.
A suitable rescheduling strategy is needed to prevent this.
Another issue is the different behavior at weekends, when more joint activities take place than on weekdays,
resulting in an increased share of trips made using the mode car as passenger.
If a mode choice model that takes availability of ride-sharing opportunities into account is used,
it can be difficult to reproduce the correct modal split without modeling explicitly these joint activities.
When modeling travel demand for a week, it is also important to account for infrequent long-distance trips.
While the share of these trips is low, the total number is not negligible.
It seems that these long-distance trips are not well covered by the destination choice model used for
the day-to-day trips, indicating the need for a long-distance trip model of infrequent events.

\end{abstract}

\section{Introduction}
	Historically travel demand modeling has been carried out for an analysis period of one day using macroscopic
	models that produce aggregate data for a day~\citep{mcnally2000fourstep}.
	With the emergence of activity-based models~\citep{bhat1999activity} and 
	agent-based models~\citep{raney2003truly} much 
	more fine-grained results were possible.
	While much progress has been made in this direction,
	the typical analysis period used in travel demand models
	like the Day-Activity-Schedule approach~\citep{bowman2001activity},
	TRANSIMS~\citep{rilett2001overview},
	MATSim~\citep{horni2016multi}, ALBATROSS~\citep{arentze2004learning}, 
	Aurora/FEATHERS~\citep{arentze2006aurora2feathers},
	or Sacsim~\citep{bradley2010sacsim}
	is still one day.

	In travel behavior research it has become common to use multi-day surveys
	to study phenomena like 
	repetition, variability of 
	travel~\citep{huff1986repetition,hanson1988systematic,pas1995intrapersonal},
	or the rhythms of daily life~\citep{axhausen2002observing}.
	\citet{jones1988significance}
	point out the importance of multi-day travel surveys for the assessment of transport policy measures,
	since one-day surveys
	do not allow assessing how severe individual persons are affected by a policy measure. 

	Taking these results of travel behavior research into account,
	travel demand models should aim to model periods longer than a day.
	Of the established travel demand models, however,
	only MATSim has at least envisioned
	an extension of the simulation period from a day to a week~\citep{ordonez2012simulating},
	followed by a prototypical implementation, which has been applied to a
	1\%~scenario~\citep{horni2012matsim}.
	Some hints indicate that Aurora might be able to simulate
	a multi-day period~\citep{arentze2010agent}, but the focus of this work
	is on the rescheduling of daily activity plans.
	A specific multi-day travel demand model has been presented 
	by~\citet{kuhnimhof2009multiday} and \citet{kuhnimhof2009measuring},
	but the implementation has only prototypical character
	and the model is only applied to a sample of 10\,000 agents.
	A model for generating multi-week activity agendas,
	which considers household activities, personal activities, and joint activities,
	is presented by \citet{arentze2009need},
	based on the theory of need-based activity generation described by~\citet{arentze2006new}.
	The model has been extended by~\citet{nijland2014multi} to take long-term planned activities
	and future events into account.
	A destination choice model that could be used in combination with the 
	activity generation model is outlined by~\citet{arentze2013location}.

	When \mobitopp was initially designed, a simulation period of 
	one week was already envisaged~\citep{schnittger2004longitudinal}.
	However, the first implementation supported only the simulation of one day.
	After some early 
	applications of mobiTopp~\citep{gringmuth2005impacts,bender2005enhancing},
	which stimulated further extensions,
	the development almost ceased and work at \mobitopp was limited to maintenance.
	Nevertheless, \mobitopp was still in use for transport planning, 
	for example in the Rhine-Neckar Metropolitan Region~\citep{kagerbauer2010mikroskopische}.
	In 2011 the \mobitopp development was resumed~\citep{mallig2013mobitopp} 
	and a simulation period of one week was finally implemented.
	Using this new feature
	\mobitopp was successfully applied to the Stuttgart Region, a metropolitan area in Germany, as study area.
	This work was the foundation for further extensions 
	in recent years,
	like 
	ridesharing~\citep{mallig2015passenger},
	electric vehicles~\citep{weiss2017assessing,mallig2016modelling},
	and carsharing~\citep{heilig2017implementation}.

	This paper describes the current state of \mobitopp. 
	The body of the paper is divided into three parts. 
	The first part (Section~\ref{sec:mobitopp}) describes the overall 
	structure of the \mobitopp framework.
	The second part (Section~\ref{sec:extensions}) describes the extensions made to \mobitopp during the last years,
	which are not part of \mobitopp's core.
	The third part (Section~\ref{sec:application}) describes an application of \mobitopp. 
	It contains the specifications of the 
	actual mode choice and destination choice models used for this scenario and the simulation results.
	The simulation results show that simply extending a single-day model 
	to a multi-day horizon already leads to useful results.
	However, there are some issues that need special attention,  
	namely ensuring realistic start times of activities, 
 	joint activities, and seldom-occurring long-distance trips.

	\section{The \mobitopp model}
	\label{sec:mobitopp}
	\mobitopp is an activity-based travel demand 
	simulation
	model 
	in the tradition of \emph{simulating activity chains} \citep{axhausen1989simulating}.
	It is based on the principle of agent based simulation~\citep{bonabeau2002agent},
	meaning that each person of the study  area is modeled as an agent. 
	An agent in this context is an entity that makes decisions autonomously, individually, 
	and situation-dependent
	and interacts with other agents.
	In \mobitopp, 
	each agent has an individual activity schedule (activity chain)
	that is executed over the simulation period,
	making
	decisions for destination choice and mode choice.
	These decisions are based on discrete choice models.
	Since \mobitopp does not yet contain an internal traffic assignment procedure 
	and relies on external tools for this purpose,
	interactions 
	between agents occur 
	in the basic version of \mobitopp
	only indirectly by
	availability  or non-availability of cars in the household context.
	When a household's last available car is used by an agent, the mode \emph{car as driver} is not longer available
	for other household members until a car 
	is returned to the household's car pool.
	The mode choice options of the other household members are therefore restricted by the action of the agent.
	Direct interaction between the agents occurs in the case of ridesharing;
	this functionality is provided by
	one of \mobitopp's extensions (see Section~\ref{sec:extensions:ridesharing}). 
	When this extension is activated,
	agents
	travelling by car offer ridesharing
	opportunities; agents that choose the mode \emph{car as passenger} actively seek for
	ridesharing opportunities. Agents travel together in the same car if a ridesharing request matches a
	corresponding offer.

	The agents' activities and trips are simulated
	chronologically over a simulation period 
	up to one week.
	The temporal resolution is one minute; the spatial resolution is based on traffic analysis zones.
	\mobitopp has been successfully applied to a study area with more than two million inhabitants 
	distributed over
	more than thousand traffic analysis zones~\citep{mallig2013mobitopp,kagerbauer2016modeling}.

\begin{figure}
		\centering
		\includegraphics[width=\textwidth]{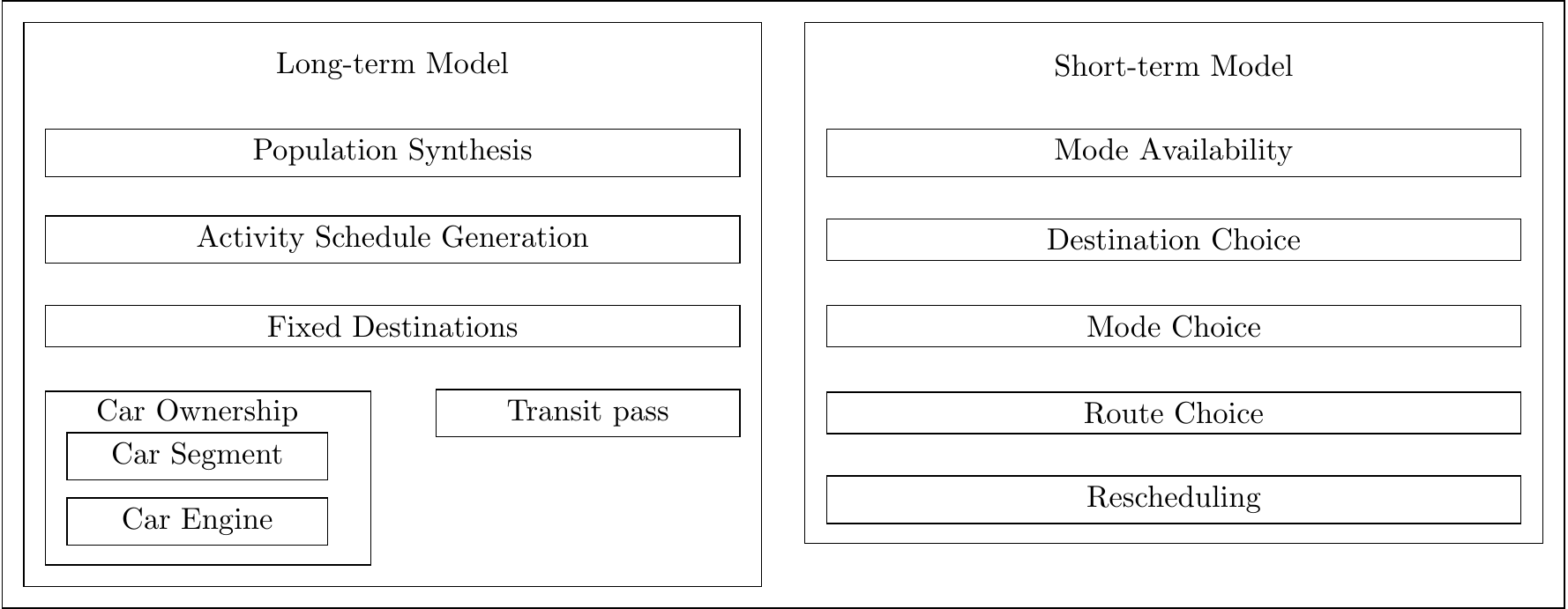}
		\caption{Structure of \mobitopp: long-term and short-term model.}
		\label{fig:mobitopp}
	\end{figure}

	\mobitopp consists of two major parts, the \emph{\stageinit} and the \emph{\stagesimu}, 
	each of them making use of several modules (see Figure~\ref{fig:mobitopp}). 
	The \stageinit represents the long-term aspects of the system like population synthesis, assignment
	of home zone and zone of workplace, car ownership, and ownership of transit pass.
	These long-term aspects define the framework conditions for the subsequent travel demand simulation.
	The \stagesimu models the travel behavior of the agents, consisting of destination choice
	and mode choice, over the course of the simulation period of one week.

	\mobitopp is implemented in JAVA using the object-oriented design paradigm.
	Every module is described by an interface.
	Typically, several implementations for each module exist, which are easily exchangeable.
	Exchanging existing implementations of a module is basically a matter of configuration.
	New implementations can be plugged-in easily.
	There is typically a default implementation, which provides only the basic functionality.
	More complex behavior is realized by specialized implementations, typically as a result of a
	specific research project.

	\subsection{Long-term model}
	The first part of the \stageinit is the population synthesis module.
	Households and persons are generated for each zone based on total numbers of households and persons
	given on the level of 
	traffic analysis zones and distributions of the households' and persons' attributes. 
	The corresponding zone is assigned as home zone.
	An activity schedule is assigned to each person.
	In addition, long-term decisions like workplace or school place, car ownership and transit pass ownership
	are modeled.
	These assignments remain fixed for the following \stagesimu.

	\subsubsection{Population synthesis}
	Central input of the population synthesis is the data of a household travel survey.
	Preferably, a survey for the planning area is used; however nationwide 
	surveys like the German Mobility Panel~\citep{wirtz2013panel} can 
	be used
	as replacement.
	The population of each zone is generated by repeated random draws of households and the corresponding persons 
	from the survey data. The distributions of households' and persons' attributes are taken into account
	by an appropriate weighting of a household's probability to be drawn.

	The population synthesis uses a two-stage process similar to the method described 
	by~\citet{mueller2011hierarchical},
	which is based on the idea of iterative proportional fitting 
	introduced by~\citet{beckman1996creating} to population synthesis.
	In the first stage,
	an initially equally distributed weight is assigned to each household.
	These weights are subsequently
	adjusted in an iterative process until the weighted distribution of the households matches the given distribution
	of the household types and the distribution of the persons weighted by the corresponding household weight matches
	the given marginal distributions of the persons' attributes.
	In the second stage,
	the corresponding number of household 
	for each household type 
	is drawn randomly
	with replacement from the weighted distribution of households. 
	The survey household acts as prototype for the household in the model, meaning that the household created in the
	model has the same attributes 
	as the prototype household. 
	Likewise,
	an agent is created 
	for each person of the survey household,
	which has
	the same attributes 
	and the same \activityprogram
	as the survey person.

	\subsubsection{Activity Schedule Generation}
	Once the agent is created, he is assigned an \activityprogram for a complete week.
	The \activityprogram consists of a sequence of activities with the attributes 
	purpose, planned start time, and duration.

	The default implementation of the \emph{\activityprogram creator} is quite simplistic:
	it basically copies the \activityprogram{}s for all persons from the persons of a matching survey household.

	A more sophisticated implementation of an \activityprogram creator, 
	which synthetically calculates the whole weekly \activityprogram{}s,
	is under development~\citep{hilgert2017modeling}.
	This \activityprogram creator, called \emph{actiTopp}, 
	consists of a hierarchy of submodels with the levels week, day, tour, and activity.
	Multinomial Logit models (MLM) 
	are used for decisions with a finite number of alternatives, for example the number
	of activities per tour. 
	Continuous variables like start time are determined using a hybrid approach. 
	A MLM is used to determine the broad period during the day; 
	a random draw from an empirical distribution is used to determine the exact time during the broad period.

	The upper level of the hierarchy models for each activity type the decisions relevant for the whole week:
	the number of days an activity of this type will take place, the available time for each activity type, 
	and the usual start time of the main tours.
	The levels below (day, tour, activity) use an approach similar to 
	the Day Activity Schedule approach~\citep{bowman2001activity}, but take also into account 
	the decisions made at the week level.
	The day level determines the main tour and the number of secondary tours.
	The tour level determines the main activity of the secondary tours and the number of secondary activities
	for each tour.
	The activity level determines the activity types for the secondary activities and the activity durations.

	\subsubsection{Assignment of workplace}
		Workplace and school place typically remain stable over a longer period
		and are
		therefore modeled in the
		\stageinit and kept constant during the \stagesimu.
		In the \stagesimu, no destination choice is made for activities of
		type work or education; instead the location assigned in the \stageinit is used.
		The assignment of workplace and school place is based on external matrices, 
		called \emph{commuting matrices} in \mobitopp,
		representing 
		the distribution of workplaces and school places for the inhabitants of each zone.
		For Germany, matrices of this type 
		can be acquired from the Federal Employment Agency for the workplace
		on the level of municipalities;
	  however, disaggregation to the level of traffic analysis 
		zones is necessary.
		Disaggregation can be done proportional to the 
		size of an appropriate variable, known for each zone,
		the number of inhabitants for the residential side,
		and for example the number of workplaces or total size of office space for the work-related side
		of each relation.
		For school places, this type of data is typically not available in Germany, so one has to resort to modeled 
		data.
		Ideally, matrices can be adopted from an existing macroscopic model.

		The agents are assigned workplaces and school places 
		based on these matrices.
		Agents whose prototypes in the survey have reported long commuting distances are assigned workplaces
		with long distances from their home zones, while agents whose prototypes reported short commuting
		distances are assigned workplaces close to their home zones.
		That way it is ensured that the commuting distance is consistent with the \activityprogram, which is taken from
		the survey data.
		The assignment of workplace and school place is done for each home zone in the following way:
		The number of workplaces for the current zone giving by the commuting matrix is normalized to match
		the total number of working persons living in the zone. The workplaces are ordered by increasing distance. 
		The working persons are ordered by increasing commuting distance reported by their prototype in the survey.
		Then each $k$th working person is assigned the $k$th workplace.
		The procedure for the assignment of school places is the same.

	\subsubsection{Car ownership}
		The number of cars, each household owns, is an attribute used by the population synthesis module
		and is therefore already defined.
		The car ownership model determines the type of car in terms of segment and engine type.
		It consists of two submodels: a car segment model and a car engine model.
		The default implementations of these model are mere placeholders for more complex models 
		that can be plugged-in as needed.
		The default car segment model always assigns a car of the midsize class.
		The car engine model assigns randomly either a combustion engine or an electric engine, while the probability
		to assign an electric engine is configurable.
		More elaborate implementations of the models were developed as part of \mobitopp's
		electric vehicle extension~\citep{weiss2017assessing}.

	\subsubsection{Transit pass}
	The transit pass model decides for each agent whether he owns a transit pass or not.
	The model is a binary logit model using the attributes sex, employment status, car availability,
	and the households' number of cars divided by household size.

	\subsection{Short-term model}
	The main components of the \stagesimu (see Fig.~\ref{fig:mobitopp})
	are the \emph{destination choice model}, \emph{the mode choice model} and the \emph{mode availability model}.
	The destination choice model and the mode choice model are used directly to model the decisions of the agents.
	The mode availability model 
	is an auxiliary model used to determine the agent's available modes.
	It is used by the mode choice model and by the destination choice model.

	During the \stagesimu, the travel behavior of all agents is simulated 
	simultaneously and chronologically. 
	The simulation starts
	at Monday 00:00 and typically ends 
	at Sunday 23:59 for a simulation period of one week.
	During the simulation period, the agents execute their \activityprogram{}s.
	Each agent typically starts the simulation performing an activity at home.
	When an agent has finished his current activity, 
	he inspects his activity schedule and identifies the next activity.
	For this activity, he 
	makes a destination choice followed by a mode choice, 
	taking into account the
	available modes. Then he makes the trip to the chosen destination using the selected mode. When he reaches
	the destination, he starts performing the next activity.

	\mobitopp uses a fixed order of destination choice first and mode choice second.
	This fixed order can be seen as a certain limitation of the model, since in reality these decisions are not
	necessarily made in this order. 
	In case of intermediate stops of a tour made by car, for example, the mode choice decision is made
	at the beginning of the tour, while the locations of the intermediate stops may be still unknown.
	Destination choice for these stops may be made at later point of time during the tour.
	We made the decision to use a fixed order because it simplifies the implementation.
	Besides, in many cases the destination choice is made before the mode choice decision,
	for example for work and school trips.
	The issue of mode choice before destination choice for intermediate stops is addressed
	by the mode availability model (see Section~\ref{sec:modeavailability}).
	This model allows a mode choice for intermediate stops only when the mode used for the tour
	is one of the modes walking, car as passenger, or public transport.
	In case of the modes car as driver or cycling, the agent is constrained to the currently used mode.
	So the order of destination choice and mode choice is virtually reversed in this case: 
	after the mode choice for the first trip of the tour only destination choices are 
	made for the trips afterwards.
	For home-based trips that do not involve an activity with fixed location
	destination choice and mode choice can be considered as simultaneous.
	The joint probability for each combination of destination and mode can be partitioned
	into a marginal probability and a conditional probability~\citep{benakiva1974structure},
	which leads to a Nested Logit structure~\citep[Ch.~10]{benakiva1987discrete_choice_analysis}.
	The applicability of this approach in \mobitopp has been shown by \citet{heilig2017largescale}.

\begin{figure}[htb]
\centering
\includegraphics[scale=0.7]{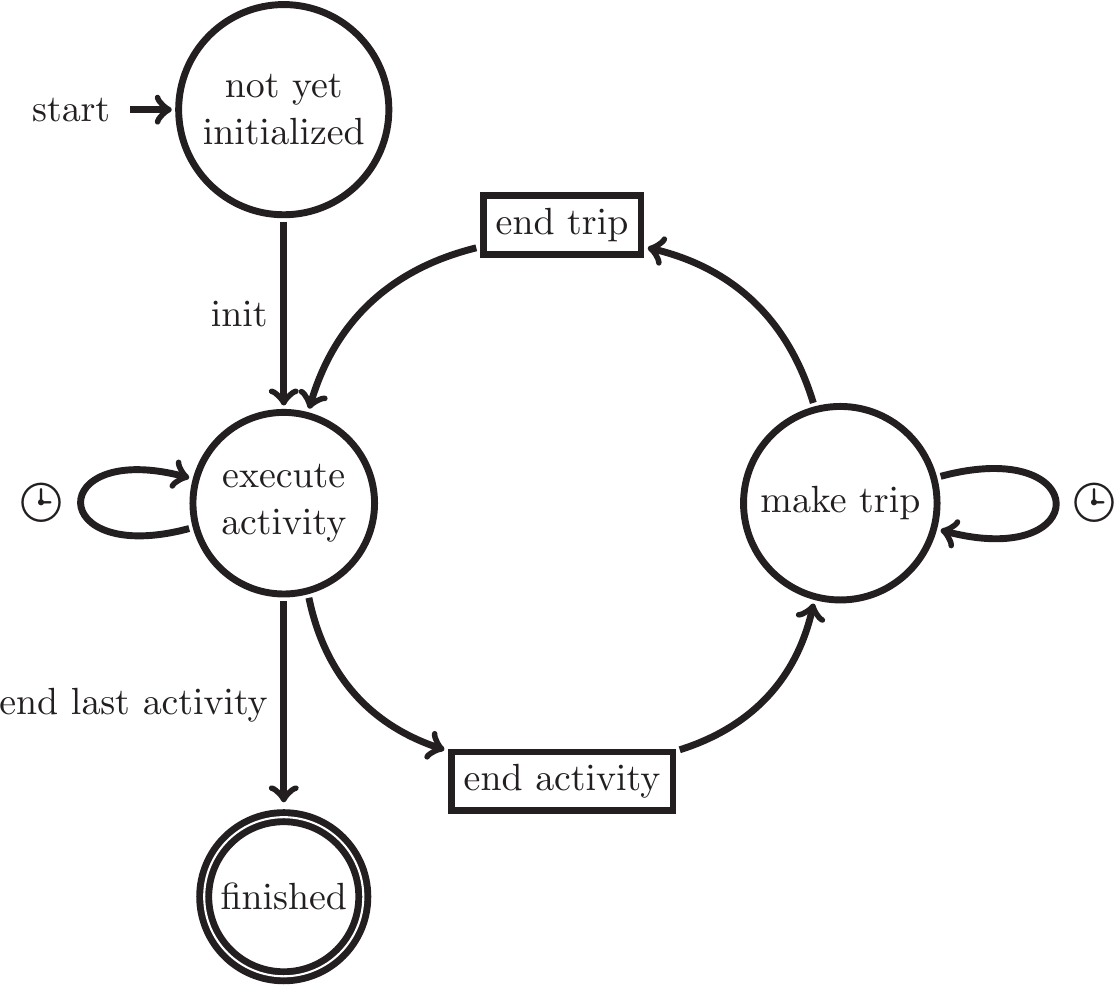}
\caption{State diagram describing the behavior of an agent during \mobitopp's \stagesimu.}
\label{fig:simulationstage}
\end{figure}

	The behavior of an agent can be described as a state diagram (Fig.~\ref{fig:simulationstage}).
	The circles denote states and the arrows denote transitions between states. The clocks at the arrows indicate
	that the agent remains in these states until some specific time has elapsed, 
	i.\,e. until the trip is finished or the activity is finished.
	
	The agent starts in the state \stateUninitialized; after initialization, the state changes to \stateActivity.
	The agent remains in this state for the duration of the activity. When the activity is finished, the state
	changes to \stateTrip. The agent remains in this state until the trip is finished and the state changes
	again to \stateActivity.
	As long as there are more activities left in the activity program,
	the cycle \stateActivity--\stateTrip--\stateActivity repeats.
	When the last activity is finished the state changes to \stateFinished.

	The agents act
	at the transitions between the states.
	At the transition from state \stateActivity to state \stateTrip, marked as \emph{end activity},
	the agents 
	makes
	first a destination choice and then a mode choice. 
	If the agent is currently 
	at home and the mode chosen is \emph{car as driver},
	the agent takes one of the household's available cars. This car is then not longer available for other agents
	until the agent that has taken the car returns home again. 
	At the transition from state \stateTrip  to state \stateActivity,
	marked as \emph{end trip},
	information about the trip made is written to the trip file.
	If the trip was a trip back home made by mode \emph{car as driver}, 
	the car is returned to the household's car pool and
	is again available to other members of the household.
	During initialization, at the transition from state \stateUninitialized to state \stateActivity,
	the first activity is initialized.
	If the activity is not of type \emph{at home}, a destination choice and a mode choice are made for 
	the trip preceding the activity.
	If the mode is \emph{car as driver} the agent is assigned a car and
	the car is removed from the pool of available cars of his household.

	\subsubsection{Destination choice}
	\label{sec:destination_choice}
	The destination choice model distinguishes between two types of activities: activities with fixed locations
	(work, school, at home) and activities with flexible locations, for example shopping or leisure.
	For activities with fixed locations, no destination choice is made, since these destinations have already
	been determined by the \stageinit.
	For activities with flexible locations, a destination choice is made 
	on the level of traffic analysis zones
	using a discrete choice model.
	Different implementations of destination choice models exist for \mobitopp.

	The different implementations of the 
	destination
	choice model have in common that they do not only consider
	travel time and cost for the trip to the potential destination, 
	but also travel time and cost for the trip to the next fixed location 
	the agent knows he will visit.

	The actual specification of the destination model used for the results shown in this paper is given in 
	Section~\ref{sec:implementation_destination_choice}.

	\subsubsection{Mode choice}
	\label{sec:mode_choice}
	In \mobitopp a trip is a journey from the location of one activity to the location
	of the next activity. 
	\mobitopp models only the main transportation mode for each trip;
	trips with several legs travelled by different modes are not supported.
	So every trip has exactly one mode.
	In its basic version,
	\mobitopp distinguishes between five modes: walking, cycling, public transport, car as driver,
	and car as passenger. 
	Two additional modes, station-based carsharing and free-floating carsharing,
	are provided by an extension~\citep{heilig2017implementation} 
	(see Section~\ref{sec:extensions:carsharing}).

	The actual available choice set is situation-dependent, consisting of a non-empty subset of the full choice set.
	\mobitopp aims at modeling the available choice set realistically. 
	This means for example that an agent, who is not at home and has arrived to 
	his current location by public transport 
	should not have available the modes cycling and car as driver for his next trip.
	Or if an agent is at home and all cars of his household are currently in use by other agents, then
 	the agent should not be able to choose the mode car as driver.
	The complexity of these dependencies is captured in the \emph{mode availability model},
	which is described in the next section.

	The choice between the available modes is made by a 
	discrete choice model, which uses
	typically service variables of the transport system like travel time and cost
	and sociodemographic variables of the agents.
	The mode is selected by a random draw
	from the resulting discrete probability distribution, 
	The actual discrete choice model used for mode choice in \mobitopp is configurable.

	The actual specification of the mode choice model used for the results of this paper is given in 
	Section~\ref{sec:implementation_mode_choice}.

	\subsubsection{Mode availability model}
	\label{sec:modeavailability}
	The actual choice set, an agent has available, depends on the current situation of the agent, 
	i.\,e. the current location, the previous mode choice 
	and the mode choices of the other agents of the same household.
	The most important factor is the agent's current location. 

	If the agent is at home, in principle all modes are
	available independently of the mode used before. However, the mode car as driver is not available if the agent
	does not hold a driving license or the household's cars are currently all in use. 
	It is assumed that every agent owns a
	bicycle, so if the agent is at home the mode cycling is always available.

	If the agent is not at home, the available choice set depends essentially on the mode used before.
	If the previous mode has been car as driver or cycling, only the mode used before is available for the next trip.
	This approach is based on the idea that a car or a bicycle that has been used at the start of a tour has
	eventually to return home. This approach is a small oversimplification, since it does not allow for tours 
	that start with the mode cycling or car as driver, are followed by a sub-tour using another mode
	that ends at the location where the vehicle has been left, and are finished by a trip using the initial mode.
	However, tours of this type are rarely found in reported behavior~\citep{kuhnimhof2009measuring}.

	If the agent is not at home and the previous mode is one of the \emph{flexible modes} walking, 
	public transport, or car as passenger,
	the choice set for the next trip consists of these three modes. 
	The modes car as driver and cycling are not available,
	since these modes require a vehicle that is typically not available if the trip before is made by another mode.

	\begin{figure}
		\centering
		\subfigure[No rescheduling.]{
			\includegraphics[width=\textwidth]{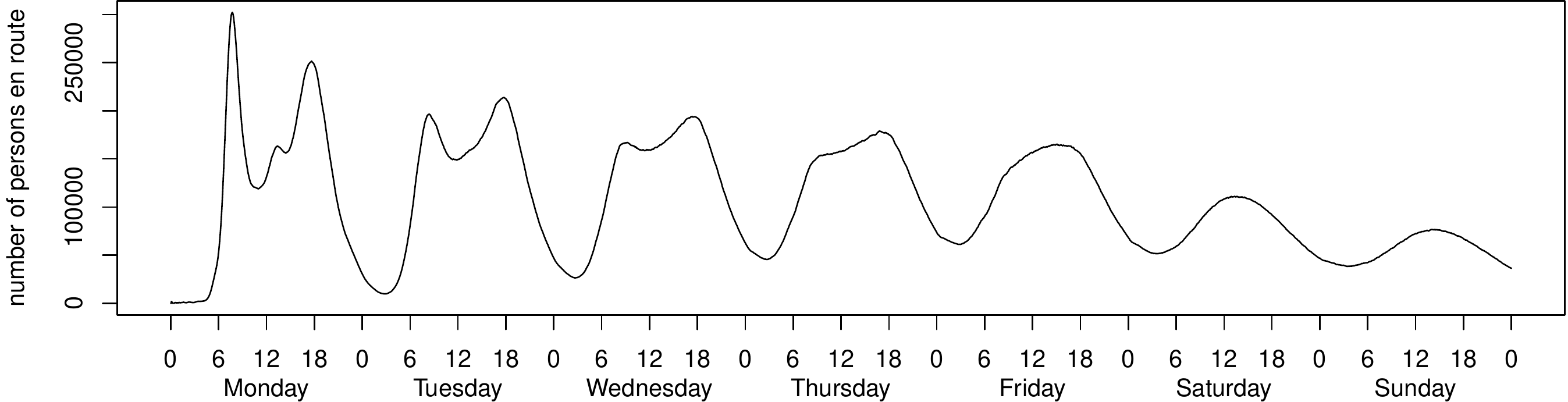}
			\label{fig:rescheduling:no}
		}
		\subfigure[Remaining activities truncated when end of day reached.]{
			\includegraphics[width=\textwidth]{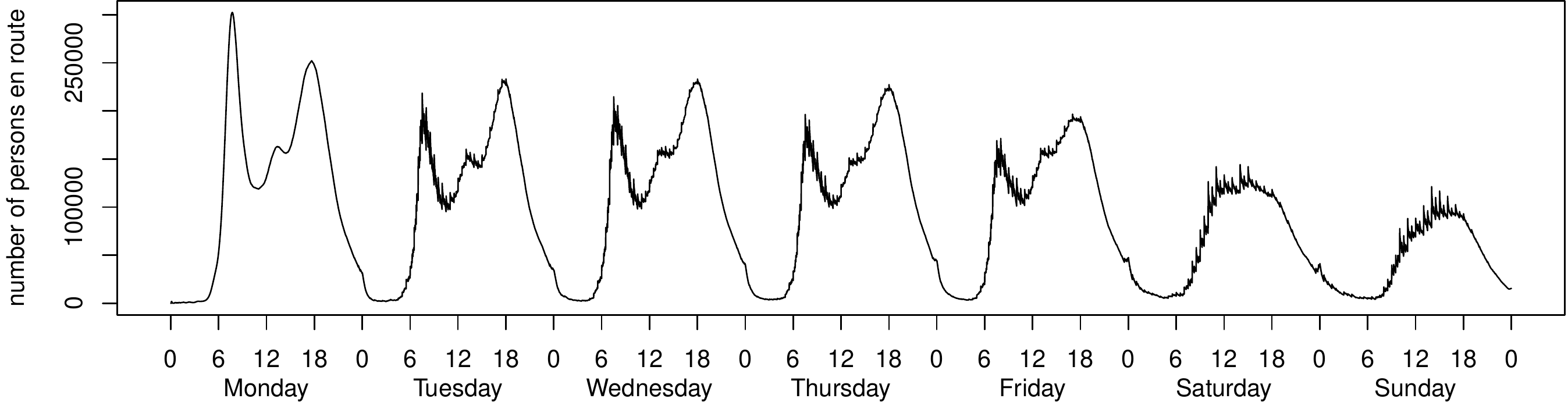}
			\label{fig:rescheduling:truncate}
		}
		\subfigure[Activities between current activity and next activity at home skipped.]{
			\includegraphics[width=\textwidth]{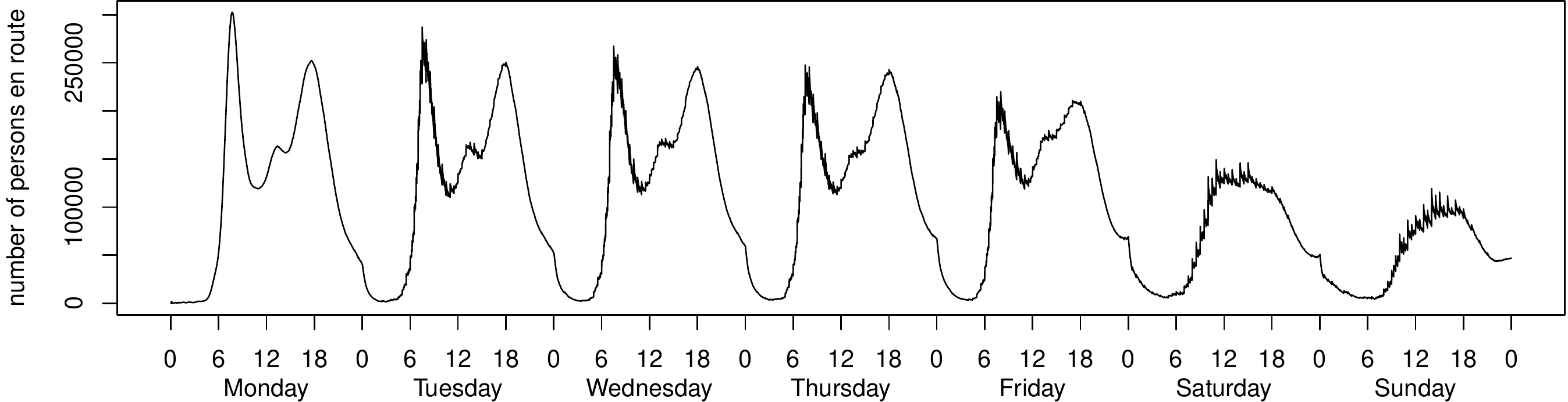}
			\label{fig:rescheduling:skip}
		}
		\caption{Times series of persons en route for different rescheduling strategies.}
		\label{fig:rescheduling}
	\end{figure}

	\subsubsection{Rescheduling}
	For each agent, the simulation flow is highly influenced by his \activityprogram.
	This \activityprogram\  consists of
	a sequence of activities for each day, with a given 
	planned start time and a duration for each activity,
	both together implying a planned end time.
	However, the travel times during the simulation may 
	not match exactly
	the gaps between 
	the planned end of an activity and the start of the next activity.
	In this case, the start time of the following activity deviates from the planned start time,
	since an agent starts his activity immediately after arriving at the location for the activity and performs the
	activity for the planned duration.
	If an agent arrives early at the destination, this means that the next activity starts earlier and
	ends earlier than planned. 
	If an agent arrives late at the destination, this means that next activity starts later and
	ends later than planned. 
	During the course of the simulation, it is possible that the start times of activities differ more and more
	form the planned start times. 
	This may finally lead to the situation that 
	some of the remaining activities of the day can not be executed
	during the day for which they have been originally scheduled.
	By this point at the latest, some rescheduling is needed.

	Several rescheduling approaches can be employed. 
	The simplest approach is just performing the activities in the planned order, ignoring the deviating schedule.
	This approach has the obvious drawback that 
	the deviations may add up in the course of the week,
	resulting in unrealistic start times for activities at later days, for example activities
	scheduled for the evening, but performed in the early morning.
	The second simplest approach is performing the activities
	until the day ends and skipping the rest. This approach has the advantage that the result of a simulation run
	with a simulation period of one day is the same as the result for the first 
	day of a multi-day simulation run.
	It has the drawback that the last activity is skipped, which is typically an \emph{at home} activity.
	For a simulation period of one day, 
	this is typically not a problem, 
	since the following \emph{at home} activity
	could take place just at the beginning of the next day. So the sequence of activities is still reasonable.
	In the case of a multi-day simulation however, this would mean that the agent does not return home 
	at the end 
	of the day, making the simulation unrealistic. 
	The third approach is keeping the last activity of the day and skipping
	activities in between the current activity and the last activity. If necessary the duration of the last activity
	can be adjusted, so that its ending time matches its planned ending time.
	A more sophisticated rescheduling approach would at first not skip any activity, 
	but try to rearrange the \activityprogram\ 
	by moving activities to other days or reducing the 
	duration of some activities~\citep{arentze2006aurora2feathers}.

	\mobitopp implements the first three approaches mentioned above.
	The third approach is enabled as default:
	activities are performed until the day ends, all remaining 
	activities of the day but the last are removed from the schedule. 
	For the last activity of a day, the duration is adjusted
	such that the first activity of the next day can start at the planned time.

	The resulting time series of the persons en route for the different rescheduling approaches 
	are shown in Figure~\ref{fig:rescheduling}.
	For Monday, all time series show a typical workday profile: a distinct morning peak
	is found around 7 a.\,m., when workers and students leave for work or school.
	Around one 1~p.\,m. is a much smaller peak, when students return home.
	The evening peak around 6~p.\,m., when people return home, is smaller but wider than the morning peak.
	In Fig.~\ref{fig:rescheduling:no},
	without explicit rescheduling, this pattern gets more and more distorted with each day.
	In Fig.~\ref{fig:rescheduling}~\subref{fig:rescheduling:truncate} and~\subref{fig:rescheduling:skip},
	with rescheduling strategies, the typical workday profile is preserved for
	from Monday to Friday. Saturday and Sunday show another profile, but as these days are not typical workdays
	a different profile is  reasonable.
	The shape of the Friday profile differs slightly from the other workday profiles, but as the Friday is at the 
	transition to the weekend this small divergence seems reasonable as well.
	The heights of the workday profiles are slightly less for Tuesday to Friday than on Monday.
	In Fig.~\ref{fig:rescheduling:skip},
	representing the rescheduling strategy where only not feasible activities before the
	home activity are skipped, the decrease in height is not so pronounced.
	The Friday profile is still lower than the Monday profile, though the total number of trips is larger on Friday.
	However, the Friday trips are distributed more evenly in time. In consequence,
	the evening peak on Friday is smaller, but broader than on Monday.
	This is the rescheduling strategy, which is enabled by defaults as it preserves the workday profile best.

	\subsubsection{Route Choice / Traffic Assignment}
	As route choice model there exists currently only a dummy implementation, returning the result of a 
	simple Dijkstra search, which is not coupled with a traffic flow model.
	So there is currently no direct feedback loop,
	where destination choice and mode choice influence travel time and the adjusted travel 
	time influences destination choice and mode choice, within \mobitopp.
	The feedback loop has to be realized using an external tool instead.
	PTV~Visum was used for this purpose since a Visum model was already available.
	This has the obvious disadvantages of leaving the agent-based world.
	We have not yet found the best option to overcome this limitation.
	One option is to use MATSim; the other option is implementing our own traffic flow simulation.

	First experiments with MATSim have shown that an integration is not straightforward since \mobitopp
	work on the level of zones while MATSim works on the level of links.
	In \mobitopp agents performing activities are located at the zone center, 
	which is connected to the road network via artificial links, so-called connectors.
	These connectors should only be used when leaving or entering a zone, but not for travel between other zones.
	When the network is converted to MATSim, the distinction between regular links and connectors is lost.
	This makes it difficult to ensure that the connectors are only used for entering or leaving a zone.
	If the capacity of the connectors is set too high and travel time too low, agents start traveling via the
	connectors and zone centers instead of the road network.
	When the capacity of the connectors is set too low, there is a lot of congestion on the connectors.
	We have recently experimented with randomly distributing \mobitopp's agent within the zone, 
	but have not yet tried out how this affects the interaction with MATSim. 
	However, according to \cite{nagel2017matsim_zone}, this approach should solve the issue.

	\section{Extensions to the base model}
	\label{sec:extensions}

	Some extensions have been made to \mobitopp during the last years, namely the implementation of 
	carsharing, 
	ridesharing,
	and electric vehicles.

	\subsection{Carsharing}
	\label{sec:extensions:carsharing}
	The \emph{carsharing extension}~\citep{heilig2017implementation} 
	extends
	\mobitopp by two additional modes: 
	station-based carsharing and free-floating carsharing.
	The \stageinit has been extended by a carsharing customer model.
	It is used to model the customer for each of the modeled car sharing companies.
	
	For the mode choice model this extension means only minor changes:
	an enriched choice set and
	different cost and increased travel time 
	compared to the mode car 
	as result of increased
	access and egress times.
	The main work for the carsharing extension regarding travel behavior is performed 
	by the mode availability model.
	Station-based carsharing cars have to be picked-up at a carsharing station, typically located in the
	vicinity of the home place, and returned to the same station.
	Thus the mode \emph{station-based carsharing} is handled like the mode car as driver:
	It is only available when the agent is at home or when it has already been used for the previous trip.
	Changing the mode is not allowed while not at home.

	The mode \emph{free-floating carsharing} is handled differently.
	Free-floating cars can be picked-up and dropped-off anywhere within a defined free-floating operating area.
	Thus free-floating carsharing is handled like the flexible modes with some restrictions.
	Free-floating carsharing is available if the zone of the current location belongs to a zone of
	the free-floating operating area and if a car is available.
	If an agent uses free-floating carsharing outside the free-floating operating area, switching the mode
	is not permitted until he returns to the free-floating operating area.

	\begin{figure}[htb]
		\centering
		\includegraphics[scale=0.7]{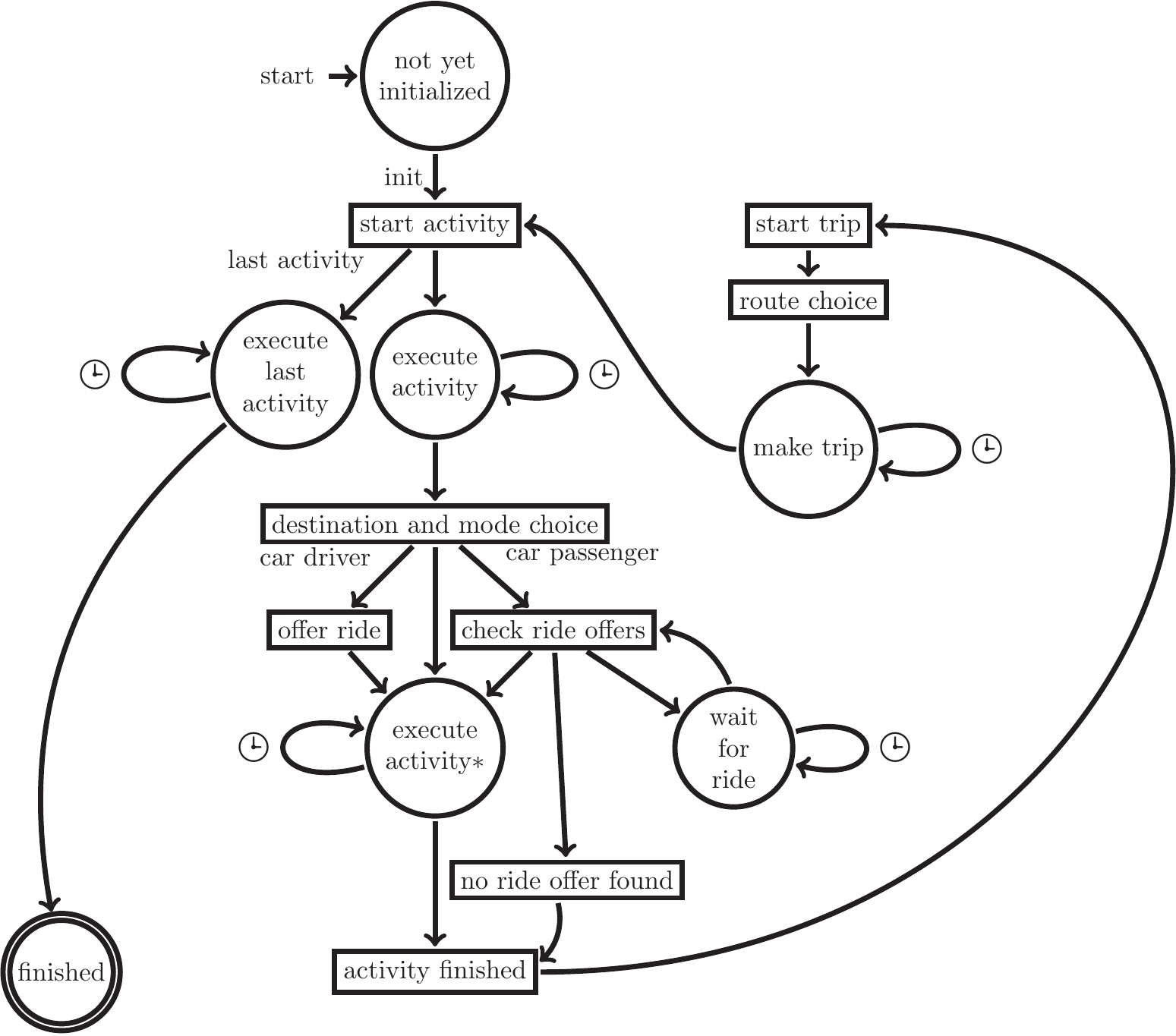}
		\caption{State diagram describing the behavior of an agent during \mobitopp's \stagesimu
			using the ridesharing extension.}
		\label{fig:simulationstage_passenger}
	\end{figure}

	\subsection{Ridesharing}
	\label{sec:extensions:ridesharing}
	The implementation of the mode \emph{car as passenger} in \mobitopp's base model is quite simplistic.
	It is assumed that this mode is always available, if not restricted to another mode.
	In reality, however, travelling as car passenger is only possible if someone else offers a ride.
	Thus the ridesharing extension of \mobitopp~\citep{mallig2015passenger} aims at modeling this mode
	more realistically.
	In this implementation, agents using a car as driver offer ridesharing opportunities. 
	Agents that have chosen the mode \emph{car as passenger} seek actively for ridesharing opportunities.

	However, when the departure time is fixed, it is practically impossible to find a ridesharing opportunity to the
	desired destination. Therefore more flexibility is needed.
	Therefore in the ridesharing extension agents make their mode choices not at the end of the activity,
	but up to 30~minutes earlier and end their activities earlier or later to match the departure time of
	the ridesharing offer.
	The extended state diagram for the ridesharing extension is shown in Figure~\ref{fig:simulationstage_passenger}.
	The state \stateActivity is split into two. Between these two states the destination choice and the
	mode choice is made.
	When the agent chooses the mode car as passenger he checks the availability of ride offers.
	If no matching ride offer is found, the agent changes in the state \stateWait where he
	periodically checks the availability of ride offers until he finds one or his maximum waiting time is elapsed.
	If the agent finds a matching ride offer or chooses one of the other modes, the next stage
	is the second part of the state \stateActivity.
	The 
	remainder
	of the diagram is in principle the same as the original state diagram.

	\subsection{Electric vehicles}
	Within the scope of the \emph{electric vehicle extension} \citep{weiss2017assessing,mallig2016modelling}
	different car classes have been created, differentiated
	by propulsion, namely combustion engine cars, battery electric vehicles, and plug-in hybrid vehicles.
	The cars have been extended to provide current fuel levels, battery levels, and odometer values.

	The \stageinit has been extended by a sophisticated car ownership model~\citep{weiss2017assessing}.
	The car ownership model consists of a car segment model and a car engine model.
	The car segment model is a multinomial logit model estimated on the data of the
	German Mobility Panel~\citep{wirtz2013panel}.
	The car engine model combines the ideas of propensity to own an electric car and suitability of an electric car.
	The propensity model models the propensity to own an electric car based on the similarity of the
	agent with early adaptors of electric cars based on sociodemographic attributes.
	The suitability model is a binary logit model 
	based on the sociodemographic attributes of the agents
	and estimated on the data generated by the CUMILE approach~\citep{chlond2014hybrid}.
	Both models calculate a probability, which are multiplied to generate a probability to own an electric car.

	In the \stagesimu, the destination choice model, and the mode choice model have been changed 
	to take the limited range into account.
	In the destination choice model, possible destinations out of range of the electric vehicle are excluded
	from the choice set if the electric car is the only mode available. A safety margin, to account for detours
	and range anxiety, is considered.
	In the mode choice model, the limited range is handled by the mode availability model.
	If the chosen destination is out of range of the electric vehicle, again taking the safety margin
	into account, the mode car as driver is not available.

	\section{Application of the model}
	\label{sec:application}
	\mobitopp has been applied to the Stuttgart Region, a metropolitan area in Germany,
	with a population of approximately 2.7~million.
	The Stuttgart Region consists of the city of Stuttgart 
	and the five surrounding administrative districts: B\"{o}blingen, Esslingen, Ludwigsburg, G\"{o}p\-pingen,
	and Rems-Murr.
	The study area is divided into 1\,012 traffic analysis zones. Additional 152 zones are used 
	to represent the surrounding land.
	The division into zones, the travel time, cost and commuting matrices,
	and the structural data have been borrowed
	from an existing macroscopic model~\citep{ptv2011verkehrsmodellierung}.

	For population synthesis, activity schedule generation,
	and parameter estimation of the different modules,
	a recent household travel survey~\citep{stuttgart2011haushaltsbefragung} has been used.
	This survey 
	was conducted between September~2009 and April~2010 in the Stuttgart Region.
	Each person, aged 6 and above, of the surveyed households was asked to fill out an activity-travel diary over
	a period of seven days. 
	The starting days of the survey period were equally distributed over all seven days of the week.
	In addition, socioeconomic data of the household and its members was gathered using a questionary.
	The sample has been generated by random draws from the municipal population registers.
	The dataset contains trip data for 275913 trips of 13731 persons in 5567 households.

	\subsection{Long-term model}
	For population synthesis and
	the generation of the \activityprogram{}s, the survey is used as input.
	The default implementation of the activity schedule generator has been used.
	The commuting matrices of the VISUM model mentioned above have been used as input for the assignment 
	of workplaces
	and school places. The default car ownership model has been used.

	\subsubsection{Transit pass model}
	Statistics on transit pass ownership was only available as totals for the whole planning area.
	The analysis of the household travel survey has shown substantially different shares of transit pass ownership
	for the different administrative districts.
	These differences could not be explained by the socio-econometric attributes alone.
	So the variable \emph{administrative district} has been added as explanatory variable.
	The model has been estimated using the method \emph{glm} contained in the \emph{stats} package
	of the statistical software package \gnuR~\citep{RCoreTeam2013R}
	based on the data of the socioeconomic questionary.
	The resulting parameter estimates are given 
	in Table~\ref{tab:transitpass}.
	As only the total number of transit passes sold was known, the sole parameter adjusted during calibration
	was the intercept, set to $0.48$.

		\begin{table}[htbp]
		\centering\small
		\begin{tabular}{@{}lQQ@{}}
				\escl{coefficient}  & \escl{estimate}  & \escl{std. error} \\
				\hline
				intercept 	  									&   0.93383  &  0.11802 \\
				female 													&  0.21752  &  0.07000 \\
				number of cars divided by household size	& -1.17629  &  0.12277 \\
				car availability&&\\
				\tabindent personal car          &  -1.32630  &  0.11780 \\
				\tabindent after consultation 		&  -0.13437  &  0.11191 \\
				employment&&\\
				\tabindent part time        			&  -0.63909  &  0.09894 \\
				\tabindent unemployed        		&  -0.56445  &  0.28009 \\
				\tabindent vocational education  &   1.44164  &  0.20866 \\
				\tabindent homemaker        			&  -1.89145  &  0.18279 \\
				\tabindent retired        				&  -0.91735  &  0.08390 \\
				\tabindent unknown        				&  -0.65950  &  0.27122 \\
				\tabindent secondary education   &   1.45025  &  0.13669 \\
				\tabindent tertiary education    &   1.87161  &  0.15488 \\
				administrative district&&\\
				\tabindent BB         						&  -1.15480  &  0.11197 \\
				\tabindent ES     					    	&  -1.30675  &  0.10157 \\
				\tabindent GP        					  &  -1.91134  &  0.16567 \\
				\tabindent LB          					&  -1.04585  &  0.09297 \\
				\tabindent WN          					&  -0.94395  &  0.09760 \\
				\hline
			\end{tabular}
		\caption{Parameter estimation results for the transit pass model.}
		\label{tab:transitpass}
		\end{table}

	\subsection{Destination choice model}
	\label{sec:implementation_destination_choice}
	The destination choice model used, for the results presented here,
	is a multinomial logit model based on the following variables: 
	purpose (type of activity), attractivity of the potential destination zone, 
	travel time and travel cost.
	Travel time and travel cost are not only based on the current zone and the potential destination zone,
	but also on the next zone where an activity with fixed location takes 
	place.

	The following utility function is used:
	\begin{equation*}
	\begin{split}
			 V_{ij} &= \beta_{time\times purpose}\cdot (t_{ij}+t_{jn}) \cdot x_{purpose} \\
							&+ \beta_{time\times employment}\cdot (t_{ij}+t_{jn}) \cdot x_{employment} \\
			 				&+ \beta_{cost\times purpose}\cdot (c_{ij}+c_{jn}) \cdot x_{purpose} \\
			 				&+ \beta_{opportunities\times purpose}\cdot \log(1+A_{j,purpose}) \cdot x_{purpose}
	\end{split}
	\end{equation*}
	where
	$V_{ij}$ is the utility for a trip from the current zone $i$ to the zone $j$.
	$A_{j,purpose}$ is the attractivity of zone $j$ for an activity of type $purpose$.
	$t_{ij}$ and $c_{ij}$ are travel time and travel cost from the current zone $i$ to zone $j$.
	$t_{jn}$ and $c_{jn}$ are travel time and travel cost from zone $j$ to the next zone $n$ of an activity with
	fixed location. 
	$x_{purpose}$ and $x_{employment}$ are dummy variables, denoting the purpose of the trip and the employment
	status of the person respectively.
	The $\beta$s are the corresponding model parameters.
	$\beta_{time\times purpose}$ and $\beta_{time\times employment}$ are the parameters for travel time, which
	vary with purpose and employment, respectively.
	$\beta_{cost\times purpose}$ is the model parameter for cost, which varies with the purpose of the trip.
	$\beta_{opportunities\times purpose}$ is the model parameter for the available opportunities,
	which also varies with purpose.

	The model parameters 
	of the destination choice model
	have been estimated
	based on the trip data of 
	the whole Stuttgart household travel survey
	using the package \emph{mlogit}~\citep{croissant2012estimation}
	of the statistical software \gnuR~\citep{RCoreTeam2013R}.
	For parameter estimation, a random sample of 100 alternatives has been used,
	as described by~\citet[Chapter 9.3]{benakiva1987discrete_choice_analysis}.
	The resulting parameter estimates are showon in Table~\ref{tab:destinationchoicemodel}.

		\begin{table}[htbp]
		\begin{minipage}[t]{0.5\textwidth}
		\centering\small
			\begin{tabular}{@{}lQQ@{}}
				\escl{coefficient}  & \escl{estimate}  & \escl{std. error} \\
				\hline
				time $\times$ purpose&&\\
					\tabindent business             &  -0.01054540 & 0.00111986\\
					\tabindent service             	&  -0.09110973 & 0.00209485\\
					\tabindent private business     &  -0.07776060 & 0.00152797\\
					\tabindent private visit        &  -0.05457433 & 0.00125026\\
					\tabindent shopping daily       &  -0.11127299 & 0.00199560\\
					\tabindent shopping other       &  -0.05947906 & 0.00171003\\
					\tabindent leisure indoor       &  -0.03639682 & 0.00141101\\
					\tabindent leisure outdoor      &  -0.06822290 & 0.00179649\\
					\tabindent leisure other        &  -0.05259948 & 0.00145475\\
					\tabindent strolling            &  -0.17635836 & 0.00357956\\
				time $\times$ employment status&&\\
					\tabindent part-time   						& -0.01155729 & 0.00074396\\
					\tabindent unemployed   					&  -0.00224253 & 0.00202341\\
					\tabindent homemaker							& -0.01375566 & 0.00103072\\
					\tabindent retired    						& -0.00343436 & 0.00062823\\
					\tabindent student    						& -0.01159740 & 0.00074088\\
					\tabindent vocational education 	&   0.00010795 & 0.00187697\\
					\tabindent other      						&  -0.00699810 & 0.00196023\\
				cost $\times$ purpose&&\\
					\tabindent business              & -0.41934536 & 0.01549222\\
				\hline
			\end{tabular}
		\end{minipage}
		\begin{minipage}[t]{0.5\textwidth}
		\centering\small
			\begin{tabular}{@{}lQQ@{}}
				\escl{coefficient}  & \escl{estimate}  & \escl{std. error} \\
				\hline
				cost $\times$ purpose (cont.)&&\\
					\tabindent service               & -0.33389025 & 0.02598859\\
					\tabindent private business     & -0.26954419 & 0.01817515\\
					\tabindent private visit        & -0.12415494 & 0.01360810\\
					\tabindent shopping daily       & -0.47753511 & 0.02667347\\
					\tabindent shopping other       & -0.33848300 & 0.02121217\\
					\tabindent leisure indoor       & -0.49940005 & 0.01844977\\
					\tabindent leisure outdoor      & -0.22842223 & 0.02129771\\
					\tabindent leisure other        & -0.31403821 & 0.01735821\\
					\tabindent strolling            & -0.70369097 & 0.04702899\\
				opportunities $\times$ purpose&&\\
				  \tabindent business            	&  0.26737545 & 0.01313610\\
				  \tabindent service             	&  0.33691243 & 0.00701132\\
					\tabindent private business     &  0.46352602 & 0.00684424\\
				  \tabindent private visit        &  0.37621891 & 0.00622737\\
					\tabindent shopping daily       &  0.27836507 & 0.00464230\\
				 	\tabindent shopping other       &  0.35537183 & 0.00467768\\
					\tabindent leisure indoor       &  0.38143512 & 0.00663500\\
				 	\tabindent leisure outdoor      &  0.29088396 & 0.00551271\\
					\tabindent leisure other        &  0.47347275 & 0.00847906\\
					\tabindent strolling            &  0.09158896 & 0.01121878\\
				\hline
			\end{tabular}
		\end{minipage}

		\caption{Parameter estimates for the destination choice model.}
		\label{tab:destinationchoicemodel}
		\end{table}

	For calibration an additional scaling parameter $\gamma = \gamma_{purpose} \cdot \gamma_{employment}$
	has been introduced, so the resulting selection probability for each zone is given by 
	\[
			P_{ij} = \exp(\gamma_{purpose} \cdot \gamma_{employment} \cdot V_{ij}) 
												/ \sum_{k} \exp(\gamma_{purpose} \cdot \gamma_{employment} \cdot V_{ik}).
	\]

		\begin{table}[htbp]
		\tiny
		\begin{minipage}[t]{0.5\textwidth}
			\begin{tabular}[t]{@{}lQQq@{}}
				\escl{coefficient}  & \esc{estimate}  & \esc{std. error} & \esc{calibration}\\
				\hline
				\multicolumn{2}{@{}l}{alternative specific constants} &&\\
				\tabindent cycling                            & -1.1356e+00 & 7.5440e-02 & \\
				\tabindent car driver                         & -1.5672e-01 & 6.5945e-02 & +0.2 \\
				\tabindent car passenger                      & -4.2941e+00 & 8.0241e-02 & -0.4 \\
				\tabindent public transport                   & -2.8372e+00 & 8.0784e-02 & +0.5 \\
				travel cost per kilometer 			           	    & -3.5730e-01 & 6.8557e-02 & \\
				travel time per kilometer      			      	  & -2.8840e-02 & 2.6580e-03 & \\
				distance in kilometer &&&\\
				\tabindent cycling        						        &  5.6343e-01 & 1.5842e-02 & \\
				\tabindent car driver                 				&  7.8261e-01 & 1.4669e-02 & \\
				\tabindent car passenger 											& 7.9470e-01	& 1.4680e-02 & \\
				\tabindent public transport 									& 7.9439e-01	& 1.4771e-02 & \\
				intrazonal trip &&&\\
				\tabindent cycling      									   	& -9.5082e-02 & 4.7767e-02 & +0.3 \\
				\tabindent car driver             						& -9.2063e-01 & 3.7774e-02 & +0.6 \\
				\tabindent car passenger       								& -1.1832e+00 & 4.4493e-02 & -0.2 \\
				\tabindent public transport              			& -2.0302e+00 & 1.1787e-01 & -0.9 \\
				\multicolumn{2}{@{}l}{transit pass} &&\\
				\tabindent cycling                 						& -4.2752e-01 & 4.1864e-02 & \\
				\tabindent car driver                  				& -8.2268e-01 & 3.3557e-02 & \\
				\tabindent car passenger            					& -1.9584e-01 & 3.6310e-02 & \\
				\tabindent public transport                   &  2.3695e+00 & 3.8847e-02 & \\
				no driving licence &&&\\
				\tabindent cycling              							& -5.9253e-01 & 8.1967e-02 & \\
				\tabindent car driver                					& -4.1309e+00 & 1.2221e-01 & \\
				\tabindent car passenger          						&  3.1978e-03 & 6.1948e-02 & \\
				\tabindent public transport                 	& -5.5511e-02 & 6.8374e-02 & \\
				female &&&\\
				\tabindent cycling                 						& -5.6703e-01 & 3.8076e-02 & \\
				\tabindent car driver                  				& -6.3911e-01 & 2.9206e-02 & \\
				\tabindent car passenger            					&  6.7420e-01 & 3.2735e-02 & \\
				\tabindent public transport                   &  8.4118e-02 & 3.6384e-02 & \\
				day of week &&&\\
				\tabindent saturday &&&\\
				\tabindent\tabindent cycling                  & -1.9626e-01 & 5.6173e-02 & +0.1\\
				\tabindent\tabindent car driver               & -4.2888e-03 & 3.6395e-02 & +0.6\\
				\tabindent\tabindent car passenger            &  5.2985e-01 & 4.1188e-02 & +0.9\\
				\tabindent\tabindent public transport         & -1.3885e-01 & 5.3039e-02 & +0.6\\
				\tabindent sunday &&&\\
				\tabindent\tabindent walking    					    & \esc{---}				 	& \esc{---}			   & -0.3\\
				\tabindent\tabindent cycling                  & -4.5429e-01 & 7.0118e-02 & \\
				\tabindent\tabindent car driver               & -5.5067e-01 & 4.6085e-02 & +1.3\\
				\tabindent\tabindent car passenger            &  1.6730e-01 & 4.9243e-02 & +1.6\\
				\tabindent\tabindent public transport         & -1.0634e+00 & 6.7111e-02 & +1.4\\[1.5ex]
				employment status &&&\\
				\tabindent vocational education &&&\\
				\tabindent\tabindent cycling        					&  1.1499e-01 & 2.0892e-01 & \\
				\tabindent\tabindent car driver         			&  5.0964e-01 & 1.5566e-01 & +0.5\\
				\tabindent\tabindent car passenger   					&  3.3904e-01 & 1.6909e-01 & \\
				\tabindent\tabindent public transport         &  4.7232e-02 & 1.7418e-01 & +0.1\\
				\tabindent infant &&&\\
				\tabindent\tabindent cycling           				& -3.6321e-01 & 7.5163e-01 & \\
				\tabindent\tabindent car driver            		& -3.7374e-02 & 1.5694e+04 & \\
				\tabindent\tabindent car passenger      			&  9.1319e-01 & 2.9921e-01 & \\
				\tabindent\tabindent public transport         &  6.9077e-01 & 6.8562e-01 & \\
				\tabindent unemployed &&&\\
				\tabindent\tabindent cycling          				& -3.1530e-01 & 1.7854e-01 & +0.1\\
				\tabindent\tabindent car driver           		& -5.8424e-01 & 1.1414e-01 & +0.3\\
				\tabindent\tabindent car passenger     				& -2.1840e-01 & 1.3849e-01 & +0.1\\
				\tabindent\tabindent public transport         &  5.8309e-02 & 1.6185e-01 & +0.1\\
				\tabindent other &&&\\
				\tabindent\tabindent cycling             			& -6.8318e-01 & 1.8341e-01 & \\
				\tabindent\tabindent car driver              	& -4.7136e-01 & 9.9071e-02 & +0.5\\
				\tabindent\tabindent car passenger        		& -1.8951e-01 & 1.2550e-01 & \\
				\tabindent\tabindent public transport         & -3.4615e-02 & 1.4701e-01 & +0.1\\
				\tabindent homemaker &&&\\
				\tabindent\tabindent cycling      						& -1.9816e-01 & 8.4113e-02 & +0.1\\
				\tabindent\tabindent car driver       				& -2.1550e-01 & 5.1162e-02 & +0.25\\
				\tabindent\tabindent car passenger 						& -2.8003e-02 & 6.5064e-02 & +0.1\\
				\tabindent\tabindent public transport        	& -1.5727e-01 & 9.4391e-02 & -0.2\\
				\tabindent partime &&&\\
				\tabindent\tabindent cycling         					&  3.2448e-01 & 5.7914e-02 & \\
				\tabindent\tabindent car driver          			&  1.3589e-01 & 4.1725e-02 & +0.2\\
				\tabindent\tabindent car passenger    				&  1.8310e-01 & 5.3525e-02 & \\
				\tabindent\tabindent public transport         &  1.0515e-02 & 5.8937e-02 & \\
				\tabindent retired &&&\\
				\tabindent\tabindent cycling          				& -6.4015e-01 & 1.0530e-01 & +0.4\\
				\tabindent\tabindent car driver           		& -6.2650e-01 & 6.6882e-02 & +0.45\\
				\tabindent\tabindent car passenger     				& -4.1904e-01 & 8.1543e-02 & +0.2\\
				\tabindent\tabindent public transport         & -7.7916e-02 & 9.6886e-02 & \\
				\tabindent student &&&\\
				\tabindent\tabindent cycling          				&  3.9692e-01 & 1.5259e-01 & +0.3\\
				\tabindent\tabindent car driver           		& -7.5693e-02 & 1.1227e-01 & +0.5\\
				\tabindent\tabindent car passenger     				&  4.6439e-02 & 1.2607e-01 & -0.2\\
				\tabindent\tabindent public transport         & -2.0048e-01 & 1.3115e-01 & -0.4\\[1.5ex]
				trip purpose &&&\\
				\tabindent business &&&\\
				\tabindent\tabindent cycling            			& -3.1571e-01 & 1.6633e-01 & -0.1\\
				\tabindent\tabindent car driver             	&  5.6801e-01 & 1.2078e-01 & +0.3\\
				\tabindent\tabindent car passenger       			&  7.4322e-01 & 1.6613e-01 & +0.8\\
				\tabindent\tabindent public transport         & -3.1422e-02 & 1.5198e-01 & +0.1\\
				\tabindent education &&&\\
				\tabindent\tabindent cycling           				& -7.2092e-01 & 7.6064e-02 & +0.05\\
				\tabindent\tabindent car driver            		& -1.0215e+00 & 8.8507e-02 & \\
				\hline
			\end{tabular}
		\end{minipage}
		\begin{minipage}[t]{0.5\textwidth}
			\begin{tabular}[t]{@{}lQQq@{}}
				\escl{coefficient}  & \esc{estimate}  & \esc{std. error} & \esc{calibration}\\
				\hline
				trip purpose (cont.)\hspace{20ex}&&&\\
				\tabindent education (cont.)&&&\\
				\tabindent\tabindent car passenger      			& -1.3862e-01 & 8.0878e-02 & +0.5\\
				\tabindent\tabindent public transport         & -1.1122e-01 & 7.9957e-02 & +0.5\\
				\tabindent home &&&\\
				\tabindent\tabindent cycling                	& -8.2356e-01 & 1.6348e-01 & \\
				\tabindent\tabindent car driver               & -6.2113e-02 & 1.0463e-01 & \\
				\tabindent\tabindent car passenger           	&  1.5310e+00 & 1.1728e-01 & \\
				\tabindent\tabindent public transport         & -1.5520e-01 & 1.2665e-01 & \\
				\tabindent leisure indoor &&&\\
				\tabindent\tabindent cycling      						& -1.6181e+00 & 1.1181e-01 & -0.3\\
				\tabindent\tabindent car driver       				& -8.4596e-01 & 6.9106e-02 & -1.2\\
				\tabindent\tabindent car passenger 						&  1.1639e+00 & 8.2267e-02 & \\
				\tabindent\tabindent public transport        	& -4.5739e-01 & 8.3763e-02 & -0.4\\
				\tabindent leisure outdoor &&&\\
				\tabindent\tabindent cycling     							& -3.1209e-01 & 7.6533e-02 & +0.1\\
				\tabindent\tabindent car driver      					&  6.0902e-01 & 6.0189e-02 & -0.2\\
				\tabindent\tabindent car passenger  					&  1.8137e+00 & 7.5388e-02 & +0.5\\
				\tabindent\tabindent public transport       	& -8.2445e-01 & 8.3885e-02 & -0.2\\
				\tabindent leisure other &&&\\
				\tabindent\tabindent cycling       						& -8.4608e-01 & 7.7222e-02 & -0.1\\
				\tabindent\tabindent car driver        				& -1.7938e-01 & 5.8599e-02 & -0.6\\
				\tabindent\tabindent car passenger  					&  1.2538e+00 & 7.4144e-02 & +0.1\\
				\tabindent\tabindent public transport         & -4.2970e-01 & 7.5348e-02 & -0.2\\
				\tabindent strolling &&&\\
				\tabindent\tabindent cycling        					& -2.3744e+00 & 1.6047e-01 & +0.5\\
				\tabindent\tabindent car driver         			& -2.9623e+00 & 1.2915e-01 & -3.0\\
				\tabindent\tabindent car passenger   					& -1.2092e+00 & 1.5164e-01 & \\
				\tabindent\tabindent public transport         & -3.6069e+00 & 2.1196e-01 & \\
				\tabindent other &&&\\
				\tabindent\tabindent cycling               		& -3.5610e-01 & 5.6294e-01 & -2.0\\
				\tabindent\tabindent car driver               &  6.6026e-01 & 3.5894e-01 & -9.0\\
				\tabindent\tabindent car passenger          	&  2.3180e+00 & 3.1746e-01 & +2.0\\
				\tabindent\tabindent public transport         &  6.7591e-01 & 3.7530e-01 & -5.0\\
				\tabindent private business &&&\\
				\tabindent\tabindent cycling    							& -7.0594e-01 & 7.1261e-02 & +0.3\\
				\tabindent\tabindent car driver     					&  2.5110e-01 & 5.2300e-02 & \\
				\tabindent\tabindent car passenger  					& 1.3480e+00  & 7.1297e-02 & +0.5\\
				\tabindent\tabindent public transport      		& -4.8652e-01 & 7.1584e-02 & +0.2\\
				\tabindent private visit &&&\\
				\tabindent\tabindent cycling       						& -8.7128e-01 & 8.2963e-02 & +0.2\\
				\tabindent\tabindent car driver        				&  4.6199e-02 & 6.0521e-02 & +0.3\\
				\tabindent\tabindent car passenger  					&  1.2351e+00 & 7.6256e-02 & +0.5\\
				\tabindent\tabindent public transport         & -1.1572e+00 & 8.4478e-02 & +0.4\\
				\tabindent service &&&\\
				\tabindent\tabindent cycling             			& -9.4805e-01 & 8.1146e-02 & \\
				\tabindent\tabindent car driver              	&  1.0449e+00 & 5.3128e-02 & +0.7\\
				\tabindent\tabindent car passenger        		&  9.8400e-01 & 8.2154e-02 & +0.3\\
				\tabindent\tabindent public transport         & -1.3799e+00 & 1.0213e-01 & \\
				\tabindent shopping daily &&&\\
				\tabindent\tabindent cycling      						& -5.6289e-01 & 6.6363e-02 & +0.3\\
				\tabindent\tabindent car driver       				&  2.5313e-01 & 5.0078e-02 & +0.2\\
				\tabindent\tabindent car passenger 						&  1.2224e+00 & 7.1302e-02 & +0.6\\
				\tabindent\tabindent public transport        	& -1.1002e+00 & 7.9112e-02 & +0.4\\
				\tabindent shopping other &&&\\
				\tabindent\tabindent cycling      						& -5.9659e-01 & 9.7271e-02 & \\
				\tabindent\tabindent car driver       				&  3.0277e-01 & 6.8813e-02 & +0.2\\
				\tabindent\tabindent car passenger 						&  1.5966e+00 & 8.5150e-02 & +0.6\\
				\tabindent\tabindent public transport        	& -2.9474e-01 & 8.7803e-02 & +0.1\\
				age &&&\\
				\tabindent from 0 to 9 &&&\\
				\tabindent\tabindent cycling                  & -1.2200e+00 & 1.9258e-01 & +0.1\\
				\tabindent\tabindent car driver               & -1.9409e+01 & 2.5083e+03 & \\
				\tabindent\tabindent car passenger            &  1.4448e+00 & 1.5028e-01 & +0.4\\
				\tabindent\tabindent public transport         & -3.6995e-01 & 1.8504e-01 & -1.2\\
				\tabindent from 10 to 17 &&&\\
				\tabindent\tabindent cycling                  &  9.0540e-01 & 1.7174e-01 & -0.2\\
				\tabindent\tabindent car driver               & -2.8867e-02 & 1.7507e-01 & \\
				\tabindent\tabindent car passenger            &  1.6410e+00 & 1.4140e-01 & +0.1\\
				\tabindent\tabindent public transport         &  6.5017e-01 & 1.4839e-01 & \\
				\tabindent from 18 to 25 &&&\\
				\tabindent\tabindent cycling                  &  6.6218e-02 & 1.4544e-01 & \\
				\tabindent\tabindent car driver               &  6.1154e-01 & 1.0340e-01 & \\
				\tabindent\tabindent car passenger            &  1.1004e+00 & 1.1831e-01 & \\
				\tabindent\tabindent public transport         &  5.1406e-01 & 1.2382e-01 & \\
				\tabindent from 26 to 35 &&&\\
				\tabindent\tabindent cycling                  & -4.8777e-01 & 7.4708e-02 & \\
				\tabindent\tabindent car driver               & -4.6501e-01 & 4.8725e-02 & \\
				\tabindent\tabindent car passenger            & -1.9761e-01 & 6.5731e-02 & \\
				\tabindent\tabindent public transport         & -8.4504e-02 & 6.7550e-02 & \\
				\tabindent from 51 to 60 &&&\\
				\tabindent\tabindent cycling                  & -2.7928e-01 & 5.6226e-02 & \\
				\tabindent\tabindent car driver               &  1.3524e-01 & 3.8045e-02 & \\
				\tabindent\tabindent car passenger            &  3.6019e-01 & 5.0391e-02 & \\
				\tabindent\tabindent public transport         &  1.8826e-01 & 5.5333e-02 & \\
				\tabindent from 61 to 70 &&&\\
				\tabindent\tabindent cycling                  & -1.2830e-01 & 9.5410e-02 & \\
				\tabindent\tabindent car driver               &  6.3149e-02 & 6.3070e-02 & \\
				\tabindent\tabindent car passenger            &  6.8302e-01 & 7.6379e-02 & \\
				\tabindent\tabindent public transport         &  3.3423e-01 & 9.0021e-02 & \\
				\tabindent 71 and above &&&\\
				\tabindent\tabindent cycling                  & -2.2630e-01 & 1.1599e-01 & \\
				\tabindent\tabindent car driver               &  2.6608e-02 & 7.2146e-02 & \\
				\tabindent\tabindent car passenger            &  6.1939e-01 & 8.7350e-02 & \\
				\tabindent\tabindent public transport         &  4.8159e-01 & 1.0401e-01 & \\
				\hline
			\end{tabular}
		\end{minipage}
		\caption{Parameter estimates for the mode choice model.
							The column \emph{calibration} contains the adjustment made as result of the calibration process.
						}
		\label{tab:modechoicemodel}
		\end{table}

	\subsection{Mode choice model}
	\label{sec:implementation_mode_choice}
	The mode choice model used is a multinomial logit model with the following utility function:
	\begin{equation*}
	\begin{split}
		V_{m} = \beta_{m,0} &+ \beta_{m,dist} \cdot x_{dist} 
												+ 	\beta_{time} \cdot x_{m,time\_km}
												+ 	\beta_{cost} \cdot x_{m,cost\_km} \\
												&+ 	\beta_{m,intra} \cdot x_{intra} \\
												&+ \beta_{m,female} \cdot x_{female}
												+ 	\beta_{m,employment} \cdot x_{employment}
												+ 	\beta_{m,age} \cdot x_{age} \\
												&+ 	\beta_{m,ticket} \cdot x_{ticket}
												+ 	\beta_{m,license} \cdot x_{license} \\
												&+ \beta_{m,purpose} \cdot x_{purpose}
												+ 	\beta_{m,day} \cdot x_{day},
	\end{split}
	\end{equation*}
	where $x_{dist}$ is the road-based distance between zone centroids, 
	$x_{m,time\_km}$ and $ x_{m,cost\_km}$ are travel time per kilometer and travel cost per kilometer 
	for the mode $m$ based on the distance~$x_{dist}$.
	The binary variable $x_{intra}$ denotes, whether the trip is an intrazonal trip, 
	whereas in addition to genuine intrazonal trips, 
	trips between zones with a distance $x_{dist} < 1\,\mathrm{km}$ are also counted as intrazonal.
	The variables $x_{female}$, $x_{employment}$, $x_{age}$, $x_{ticket}$ und $x_{license}$
	are person specific variables for sex, employment status, age group, transit pass ownership, and holding
	of a driving license. 
	The variables $x_{purpose}$ and $x_{day}$ denote the purpose of the trip (activity type) 
	and the day of the week.
	The variable day of the week has the possible values workday, Saturday, and Sunday.

	The coefficients $\beta_{time}$ and $\beta_{cost}$ are generic, i.\,e. not mode-dependant.
	The constants $\beta_{m,0}$ and the other coefficients are alternative-specific, which means they are different
	for each mode $m$.

	The model parameters 
	have been estimated using \emph{mlogit}/\gnuR\ 
	based on the trip data of
	the whole Stuttgart household travel survey.
	The resulting parameter estimates are shown in Table~\ref{tab:modechoicemodel}.

	\subsection{Traffic assignment}

	As \mobitopp does not contain a traffic flow simulation, the feedback look is realized with the help of external
	tools.
	The resulting trip file of the \stagesimu is aggregated
	to OD-matrices, differentiated by mode and time period.
	A traffic assignment is made for each period using an external
	tool and the resulting travel time matrices are fed back into the next run of the \stagesimu.

	For traffic assignment, PTV Visum has been used.
	For this purpose, all trips starting within the same hour have been aggregated into hourly OD matrices.
	Hourly matrices have been generated for each day.
	Traffic assignment has been done for 24 hourly matrices for workday, Saturday, and Sunday each.
	The hourly matrices of Tuesday have been considered as representative for a workday.
	PTV Visum's \emph{Equilibrium Lohse} procedure with 50~iterations has been used for traffic assignment.
	With this procedure, traffic assignment took about 25~minutes for each of the 72~hourly matrices.

		\begin{table}[htbp]
		\small
		\begin{minipage}[t]{0.40\textwidth}
		\centering
		\hspace{\fill}
		\begin{tabular}{@{}lD{.}{.}{2}@{}}
				$\gamma_{purpose}$ & \multicolumn{1}{r@{}}{value} \\
				\hline
				business 					& 1.6\\
				service 					& 0.95\\
				private business 	& 0.95\\
				private visit 		& 1.05\\
				shopping daily 		& 0.85\\
				shopping other 		& 0.95\\
				leisure indoor 		& 1.15\\
				leisure outdoor 	& 0.95\\
				leisure other 		& 1.05\\
				strolling 				& 3.0\\
				\hline
		\end{tabular}
		\end{minipage}
		\hspace{0.05\textwidth}
		\begin{minipage}[t]{0.40\textwidth}
		\centering
		\begin{tabular}{@{}lD{.}{.}{2}@{}}
				$\gamma_{employment}$ & \multicolumn{1}{r@{}}{value} \\
				\hline
				fulltime 										& 1.3\\
				part-time 										& 1.3\\
				unemployed 									& 0.9\\
				homemaker 									& 0.95\\
				retired 										& 1.0\\
				student primary 						& 0.75\\
				student secondary 					& 0.8\\
				student tertiary 						& 0.9\\
				vocational education 				& 1.4\\
				other 											& 1.0\\
				\hline
		\end{tabular}
		\end{minipage}
		\hspace{\fill}

		\caption{Scaling parameters of the destination choice model obtained by calibration.}
		\label{tab:destinationchoicemodel_scale}
		\end{table}

	\begin{figure}[htbp]
		\includegraphics[width=0.9\textwidth]{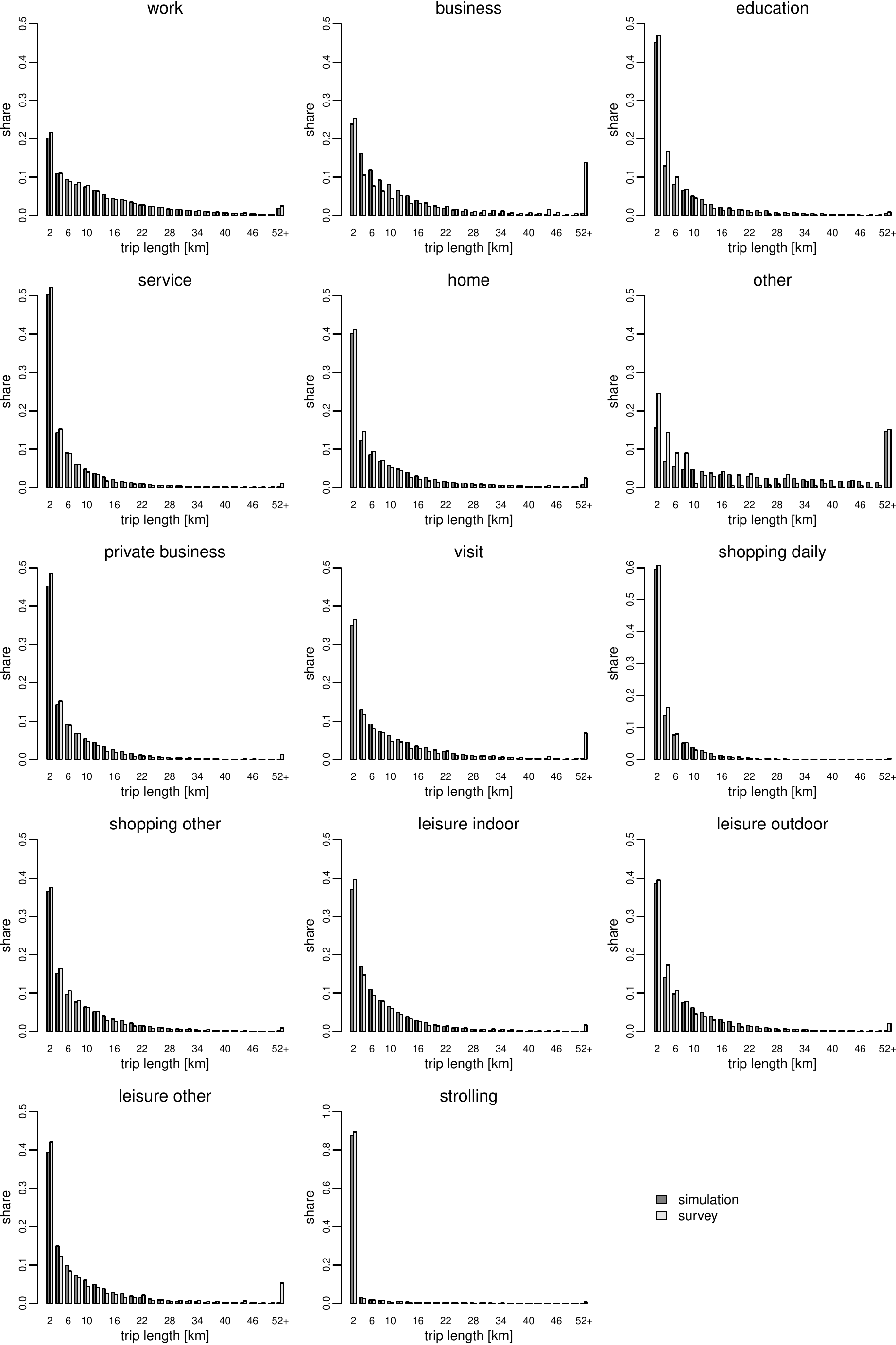}
		\caption{Trip length distribution by purpose}
		\label{result:triplength_purpose}
	\end{figure}

	\begin{figure}[htbp]
		\centering
		\includegraphics[width=0.95\textwidth]{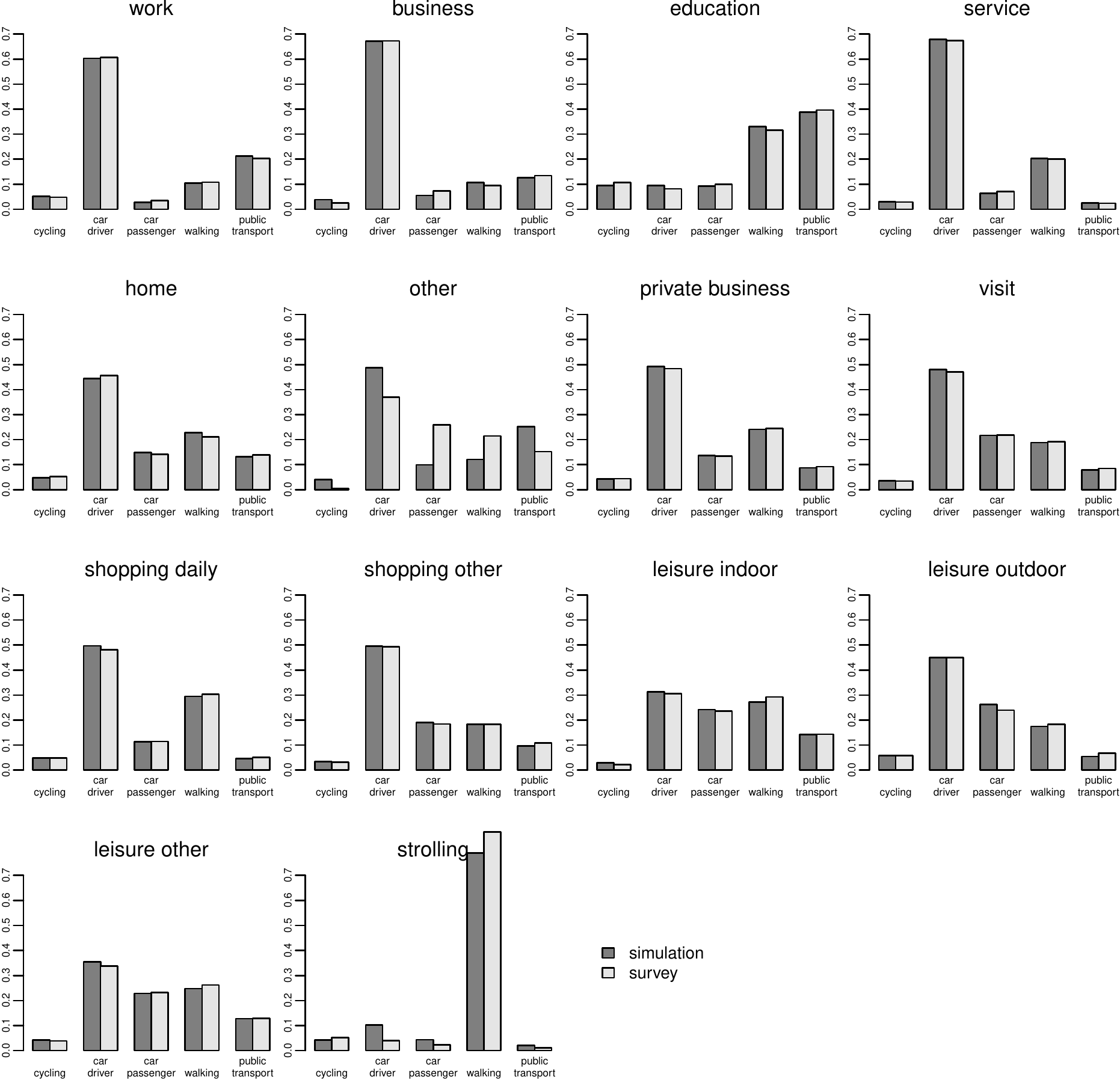}
		\caption{Modal split by purpose}
		\label{result:modalsplit_purpose}
	\end{figure}

	\begin{figure}[htbp]
		\centering
		\includegraphics[width=0.95\textwidth]{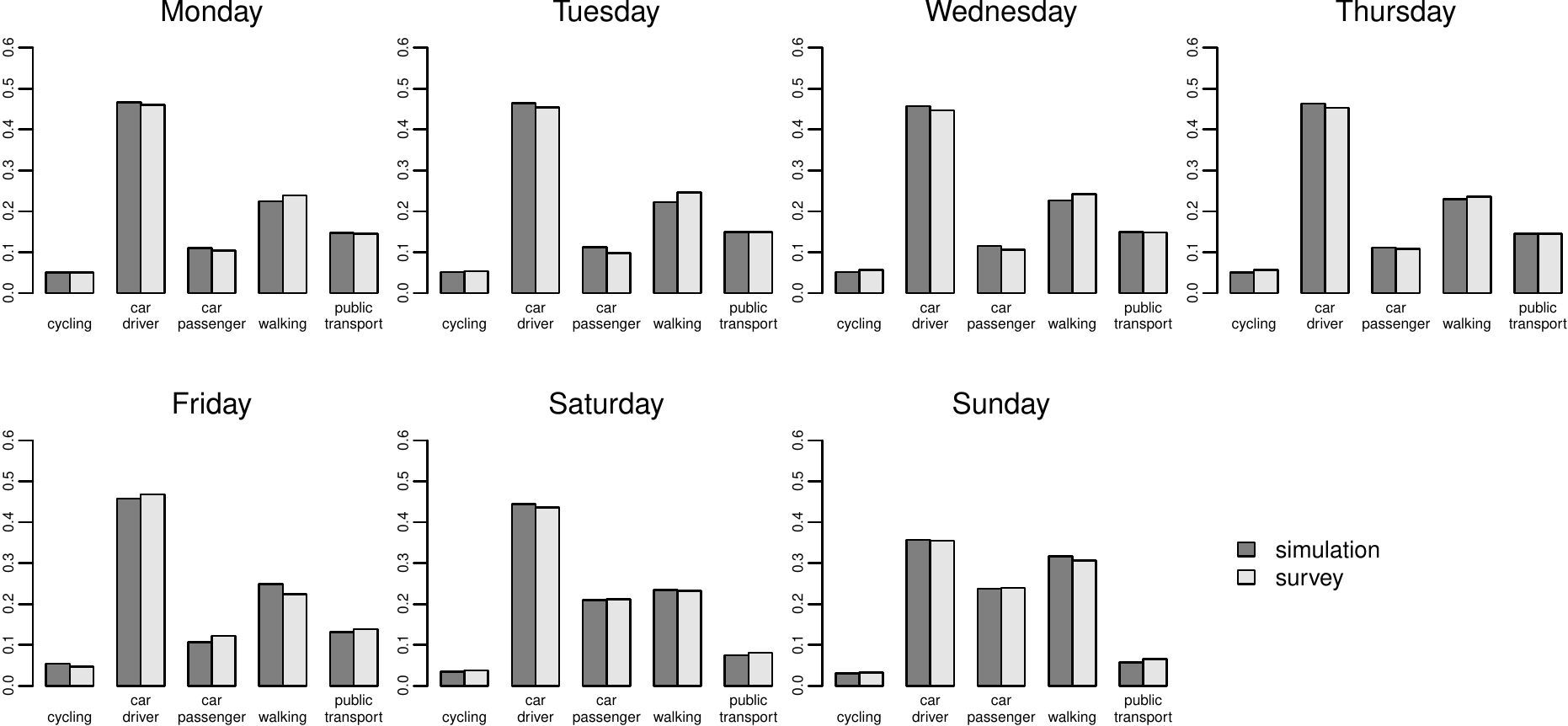}
		\caption{Modal split by day of the week}
		\label{result:modalsplit_day}
	\end{figure}

	\begin{figure}[htbp]
		\centering
		\includegraphics[width=0.95\textwidth]{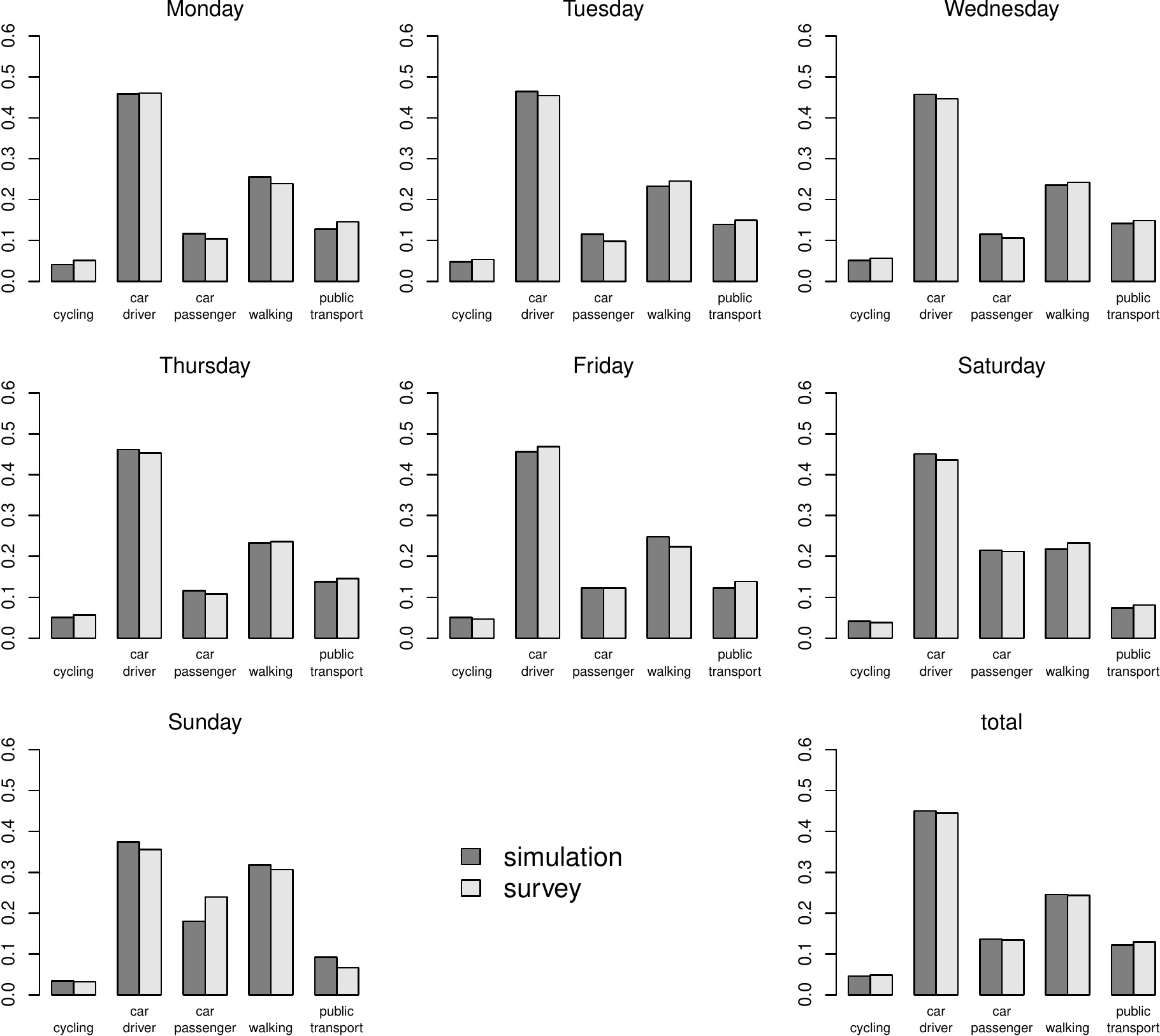}
		\caption{Modal split by day of the week using the \emph{ridesharing} extension.}
		\label{result:modalsplit_day_ridesharing}
	\end{figure}
	\subsection{Calibration}
	The destination choice model and the mode choice model have been calibrated using the data of the household
	travel survey.
	As starting point for the calibration, the estimated logit parameters 
	have been used. 
	Since the parameters of the destination choice model and the mode choices have been estimated independently,
	but actually destination choice and mode choice mutually influence each other, the initial match between
	simulated results and survey results was only moderately good,
	so subsequent calibration was necessary.

	The calibration has been based on plots of trips length distributions and modal split differentiated by several
	criteria that contrast the simulation results with the survey results, see for example 
	Figures~\ref{result:triplength_purpose} and Figure~\ref{result:modalsplit_purpose}.
	An iterative process consisting of visually judging the plots, adjusting the model parameter, and rerunning the
	simulation has been employed.
	When a good fit of the model had been reached, a new traffic assignment run was made. 
	The resulting travel time matrices were used for the next calibration round.
	Each calibration round consisted of several simulation runs with $1\%$ samples of the population until
	a good fit of the distributions was reached. These were followed by a simulation run with the whole population.
	These results were used for the next round of traffic assignment.
	In total, three complete iterations involving traffic assignment were made.

	The resulting scaling parameters for the destination choice model 
	are shown in Table~\ref{tab:destinationchoicemodel_scale}.
	The adjustments to the mode choice parameters are shown in the last column of Table~\ref{tab:modechoicemodel}.

	\subsection{Simulation results}
	\label{sec:results}
	The simulation results in
	in Figures~\ref{result:triplength_purpose} to \ref{result:modalsplit_day}
	show that the model is 
	mostly
	calibrated well regarding trip length distribution and modal split.
	The plot of the trip length distributions in Figure~\ref{result:triplength_purpose} 
	shows a good match between simulated data and survey data; 
	however the number of trips with a distance over 50~km is somewhat underestimated.
	Plots of trip length distributions by employment type (not shown here) show similar results.
	It seems that this discrepancy can be at least partially attributed to
	infrequently occurring long-distance trips like business trips, holiday trips or trips 
	to visit friends or relatives. So as future improvement, a separate model for such seldom-occurring events might
	be useful.

	The modal split by purpose in Figure~\ref{result:modalsplit_purpose} 
	shows a good match between simulated and empirical data.
	The same holds for the modal split by employment status (not shown here).
	And finally the modal split by day of the week 
	shows a good match as well (see Fig.~\ref{result:modalsplit_day}).

	Enabling the \emph{ridesharing extension}, similar results for the modal split by day could be produced
	for all days except Sunday (Fig. \ref{result:modalsplit_day_ridesharing}).
	Even an excessive adjustment of the mode choice parameters was not able to produce a higher share
	of the mode car as passenger on Sunday.
	We suppose that joint activities are the reason for this.
	Joint activities occur particularly at weekends and are combined with joined trips.
	Many of these trips are made as a combination of the modes car as driver and car as passenger.
	As \mobitopp does not yet support joint activities, 
	agents choose their destinations independently of each other 
	and therefore choose in many cases different destinations.
	Hence a joint trip is not possible and 
	this opportunity for trips as car passenger is missing.
	As we were not able to correctly reproduce the modal split for Sunday, 
	the ridesharing extension is not enabled
	as default option and and the simplistic implementation is used instead.

	\subsection{Transferability}
	All recent developments of \mobitopp are based on the model for the Stuttgart area.
	However, single-day versions of \mobitopp have been previously applied to
	the Rhine-Neckar Metropolitan Region~\citep{kagerbauer2010mikroskopische} and the Frankfurt Rhine-Main area.

	An application of the current multi-day version of \mobitopp to the Karlsruhe Region is planned.
	It is intended to use the division into zones and the commuting matrices from an existing VISUM model.
	A synthetic population will be generated based on statistics at the zonal level.
	Recalibration of the destination choice and mode choice models in order to match trip length distributions
	and modal split of the latest household travel survey conducted 
	in the Karlsruhe area~\citep{heilig2017implementation} will be necessary.

	\section{Conclusions and outlook}
		\label{sec:discussion}
	We have presented the 
	travel demand 
	simulation
	model 
	\mobitopp, which
	is able to simulate travel demand 
	over a period of one week for a large-scale study area common in practice.
	The model is derived as an extension of a single-day model; the activity schedule is extended from one
	day to a week; 
	some measures had to be taken to prevent that actual start times of activities differ too much from 
	planned start times.
	The model has been successfully calibrated: trip length distributions and modal split by several aggregation
	levels match the results of a household travel survey well; even the modal split by day of the week shows
	a good match.

	Using the model in different projects we have gained the following insights:
	First, with an increasing simulation period,  
	it is more and more important to assure that the start times of activities are correct.
	An analysis of persons en route showed that the simplistic model of simulating the sequence of
	trips and activities without further checking the start times of activities
	results in unrealistic time series.
	While for a simulation period of one day, this is not a real problem, 
	the time series is already severely blurred for the second day.
	However a very simple rescheduling strategy can already fix this issue.
	Second, as weekend travel is different from weekday travel, models or extension of models that work
	for weekdays quite well may have problems at weekends.
	In our case, the ridesharing extension was not able to reproduce the modal split at Sundays correctly
	and revealed that the activity model is not sophisticated enough for a realistic modeling
	of car passenger trips.
	So an explicit joint activity and joint trip modeling is necessary
	Third, there are infrequent extraordinary activities, which should be treated differently than the
	everyday activities of the same type.
	Comparing the simulation results with the survey data using trip length distributions
	revealed a non-negligible number of long-distance trips in the survey data for which
	the distance was not correctly reproduced by the model.
	These trips are mainly of the types \emph{business}, \emph{visit}, or \emph{leisure other}.
	As the relative number of these trips is small, we assume that these are infrequent
	extraordinary activities that have few in common with the everyday activities of the same type.
	Due to the small relative number of these activities, the destination choice model is 
	mainly calibrated for the everyday activities.
	It seems that a long-distance trip model of infrequent events is needed.

	The issue of the diverging start times of the activities has already resolved by the re\-scheduling strategy
	described in this work. However, this strategy is still rough.
	A superior solution would include the adjustment of the durations of the activities and take 
	time-space-constraints~\citep{hagerstraand1970people} into account during destination choice.
	Taking time-space-constraints into account and restricting the choice set to those locations that can be reached
	within the timespan available between two activities would prevent activities from starting later than planned.
	An early start of activities could be prevented by extending the duration of the previous activity.

	Joint activities need a flag indicating whether an activity is a joint activity or not and the information
	about the participating persons. 
	The destination choice model has to consider that only one choice is made for a joint activity and
	this choice applies for all involved persons.
	The subsequent mode choice has to be aware of the joint activity, with a high probability of a joint trip
	as result.
	As a first step, joint activities in the household context should be sufficient, already providing a considerable
	improvement. Modeling joint activities completely would require a representation of the social network,
	whereby joint activities could take place within every clique of the social network.
	The necessary information for joint activities will be eventually provided by 
	the \emph{actiTopp} implementation of the activity schedule creator module.
	The issue with the infrequent long-distance events can be resolved by subdividing the
	affected activity types
	each into \emph{distant} and \emph{local} and estimating new parameters of the destination choice model 
	for the new activity types.

\bibliographystyle{apalike}
\bibliography{references}

\end{document}